\documentclass[12pt,letterpaper,superscriptaddress]{JHEP3}
\usepackage{amssymb,amsmath,amsfonts,amstext,graphics,mathrsfs}
\usepackage{cite}
\usepackage{graphicx}
\usepackage{epstopdf}
\input epsf.tex
\usepackage{slashed}

\newdimen\tableauside\tableauside=1.0ex
\newdimen\tableaurule\tableaurule=0.4pt
\newdimen\tableaustep
\def\phantomhrule#1{\hbox{\vbox to0pt{\hrule height\tableaurule width#1\vss}}}
\def\phantomvrule#1{\vbox{\hbox to0pt{\vrule width\tableaurule height#1\hss}}}
\def\sqr{\vbox{%
  \phantomhrule\tableaustep
  \hbox{\phantomvrule\tableaustep\kern\tableaustep\phantomvrule\tableaustep}%
  \hbox{\vbox{\phantomhrule\tableauside}\kern-\tableaurule}}}
\def\squares#1{\hbox{\count0=#1\noindent\loop\sqr
  \advance\count0 by-1 \ifnum\count0>0\repeat}}
\def\tableau#1{\vcenter{\offinterlineskip
  \tableaustep=\tableauside\advance\tableaustep by-\tableaurule
  \kern\normallineskip\hbox
    {\kern\normallineskip\vbox
      {\gettableau#1 0 }%
     \kern\normallineskip\kern\tableaurule}%
  \kern\normallineskip\kern\tableaurule}}
\def\gettableau#1 {\ifnum#1=0\let\next=\null\else
  \squares{#1}\let\next=\gettableau\fi\next}

\newcommand{\gsim}{\lower.7ex\hbox{$\;\stackrel{\textstyle>}{\sim}\;$}}
\newcommand{\lsim}{\lower.7ex\hbox{$\;\stackrel{\textstyle<}{\sim}\;$}}
\def\OO{{\cal O}}
\def\DD{{\cal D}}
\def\PP{{\cal P}}
\def\LL{{\cal L}}

\def\JJ{{\cal J}}
\def\PP{{\cal P}}

\def\TT{{\cal T}}
\def\WW{{\cal W}}
\def\VV{{\cal V}}
\def\HH{{\mathcal{ H}}}
\def\SS{{\cal S}}
\newcommand{\bea}{\begin{eqnarray}}
\newcommand{\eea}{\end{eqnarray}}
\def\HHH{{\mathscr{ H}}}
\def\VVV{{\mathscr{ V}}}

\def\vv{{\mathsf{V}}}
\def\aa{{\mathsf{A}}}

\newcommand{\TeV}{\,\mathrm{TeV}}
\newcommand{\GeV}{\,\mathrm{GeV}}
\newcommand{\MeV}{\,\mathrm{MeV}}
\newcommand{\keV}{\,\mathrm{keV}}
\newcommand{\eV}{\,\mathrm{eV}}

\newcommand{\half}{{\frac{1}{2}  }}

\newcommand{\hc}{\text{ h.c. }}

\newcommand{\sld}{\partial\hspace{-0.155in}\not\hspace{0.075in}}

\newcommand{\bef}{\begin{figure}[htbp]\begin{center}}
\newcommand{\eef}{\end{center}\end{figure}}
\newcommand{\ket}[1]{|#1\rangle}

\title{
\begin{flushright}
\mbox{\normalsize SLAC-PUB-14253}
\end{flushright}
\vskip 15 pt
Nearly Supersymmetric Dark Atoms}
\author{  Siavosh R. Behbahani $^{1,2}$, Martin Jankowiak$^{1,2}$, Tomas Rube$^{2}$, Jay G. Wacker$^{1,2}$\\
$^1$Theory Group, SLAC National Accelerator Laboratory,  Menlo Park, CA 94025\\
$^2$Stanford Institute for Theoretical Physics, Stanford University, Stanford, CA 94305
}
\abstract{
Theories of dark matter that support bound states are an intriguing possibility for the identity of the missing mass of the Universe.  This article proposes a class of models of supersymmetric composite dark matter where the interactions with the Standard Model communicate supersymmetry breaking to the dark sector.  In these models supersymmetry breaking can be treated as a perturbation on the spectrum of bound states.  Using a general formalism, the spectrum with leading supersymmetry
effects is computed without specifying the details of the binding dynamics.
The interactions of the composite states with the Standard Model are computed
and several benchmark models are described.  General features of non-relativistic supersymmetric bound
states are emphasized.
}

\begin{document}

\section{Introduction}
\label{Sec: Introduction}

The nature of dark matter is unknown and its relation to the Standard Model (SM) is an open question.   The recent spate of anomalies 
in direct detection experiments \cite{DAMA} and cosmic ray signatures \cite{anomalies} has motivated re-examining the standard assumptions about the identity of dark matter.  Most models of dark matter assume that dark matter is an elementary particle with no relevant or long range interactions.  If supersymmetry is present in these models, the supersymmetric mass splittings are so large that the supersymmetric structure of dark matter is unimportant.  This article provides a framework to illustrate the exact opposite case:  dark matter is composite with long range interactions and supersymmetry breaking effects are small. 

Recent anomalies have several common features that motivate considering dark sectors that support bound states.  Bound
states naturally enjoy a hierarchy of different scales.  
Inelastic Dark Matter explanations of DAMA, {\em e.g.} \cite{iDM,StatusiDM,TheoriesOfiDM}, require several scales to reconcile the anomalies with the null results of other direct detection experiments.  
A hierarchy of scales is also employed in the Exciting Dark Matter scenario \cite{Finkbeiner:2007kk} to explain 
the 511 keV signal from INTEGRAL/SPI.  Additional structure in the dark sector is also motivated by positron excesses in cosmic ray data, which might be a result of cascade decays in the dark sector.
Examining the Standard Model, one finds a variety of different bound state systems: mesons and baryons, nuclei, atoms, 
and molecules.    
Given the prevalence of bound states in
Standard Model systems, it is natural to explore the possibility \cite{CompositeDarkMatter,Alves:2009nf,Kaplan:2009de} that dark matter 
is composed of bound states in a separate sector.

Fermions with gauge interactions are a ubiquitous ingredient in theories beyond the Standard Model. It is plausible that there are additional gauge sectors that SM fermions are not charged under. If there are no SM particles directly charged under the new gauge interaction, then experimental limits on decoupled gauge sectors are extremely weak.    
If supersymmetry breaking is only weakly mediated to the dark sector, perhaps through dark matter's interactions with the Standard Model, then the magnitude of supersymmetry breaking effects can be extremely small.  This allows for the possibility that dark matter is nearly supersymmetric.  If there are any bound states in the dark sector, the spectrum will exhibit near Bose-Fermi degeneracy.  Such weakly
coupled hidden sectors also naturally sit near the GeV scale, which makes for interesting dark matter phenomenology \cite{GevScale}
and experimental signatures \cite{ExperSignatures}.

Investigating nearly supersymmetric bound states arising from perturbative Coulombic interactions is a relatively intricate process and the standard techniques from quantum mechanics involve computing first and second order $\TT$-matrix elements and then diagonalizing the Hamiltonian.  At each step the calculation is not supersymmetric, although the final answer is supersymmetric.  Ultimately, the states have organized themselves into supersymmetric multiplets and the admixtures of different supersymmetric particles that each composite state consists of is known.   For instance, a spin zero fermion-fermion bound state will mix with a spin zero scalar-scalar bound state.   
Since phenomenological applications depend on these admixtures, it would be convenient to understand their structure and
how they generalize to other bound state systems.  Similarly, phenomenological studies would be made easier by understanding
how bound state interactions are constrained by supersymmetry.  This article develops a simple formalism to
do this using off shell superfields.

The organization of the paper is as follows.  Sec.\;\ref{Sec: NR} reviews non-relativistic supersymmetric bound states, focusing on how supersymmetry organizes the spectrum and superspin wavefunctions of the states.  The free effective action is also introduced,
which will form the basis for computing the supersymmetric interactions of the bound states.  Sec.\;\ref{Sec: SSB Effects} incorporates the effects of supersymmetry breaking into the spectrum for the case where the dominant source of supersymmetry breaking is the soft masses of the scalar constituents.  
Sec.\;\ref{Sec: Interactions} computes the interactions of the bound states when interacting with weakly coupled external gauge interactions.
Sec.\;\ref{Sec: Nearly Susy} constructs a realistic model of nearly supersymmetric atomic dark matter.   Sec.\;\ref{Sec: Discussion} discusses possible directions of future research for models along these lines, including recombination and the formation of supersymmetric molecules.
Sec.\;\ref{Sec: Conc} makes some concluding remarks.

\section{Non-relativistic Supersymmetric Bound States}
\label{Sec: NR}

This section studies how non-relativistic supersymmetric bound states organize themselves into supermultiplets.  Sec.~\ref{Sec: Wavefunctions Fields}  outlines a general procedure for determining the composition of non-relativistic bound states formed from massive superfields.  When applicable, this procedure has the advantage of sidestepping a detailed perturbative calculation in favor of some superfield algebra. This procedure is illustrated in the particular case of bound states formed from two chiral multiplets. Sec.~\ref{Sec: Effective Action} continues the study of this particular example by introducing an effective field theory description of the ground state.   This will provide the basis for Sec.~\ref{Sec: Interactions}, in which bound state interactions are discussed.

\subsection{Wavefunctions From Superspace}
\label{Sec: Wavefunctions Fields}

Non-relativistic bound states have a structure that is simple to understand because they benefit from a
good expansion parameter: the velocity $v$.  This is especially the case for two-body systems, where an expansion in powers of $v$ 
not only helps to organize calculations but also determines the relevant scales of the problem.  The gross structure
of the spectrum can be organized into principle excitations split by energies of order 
\begin{eqnarray}
m_{\text{prin}} \propto  \mu\, v^2,
\end{eqnarray}
where $\mu$ is the reduced mass.
Fine structure effects are the next order correction in the non-relativistic expansion, appearing as 
\begin{eqnarray}
m_{\text{FS}} \propto \mu\, v^4 .
\end{eqnarray}
Recent papers \cite{Rube:2009yc,Herzog:2009fw} have computed the fine structure of 
supersymmetric hydrogen through explicit calculation.
This section rederives these results by considering how supersymmetry acts on the bound states in the non-relativistic limit.
The organization of the spectrum into supermultiplets does not depend on the details of the binding dynamics except for specific quantities, such as energy splittings.  Most notably, the superspin wavefunctions are completely determined by supersymmetry alone if there are no accidental degeneracies in the spectrum.
This method of using supersymmetry to fix the superspin wavefunctions is applicable to a wider class of 
non-relativistic bound states than Coulombic bound states
and more cleanly delineates which quantities depend upon dynamics versus the structure of supersymmetry. 

For simplicity, assume that the bound state is supported by a central potential that is spin-independent at $\OO(v^2)$.  
This is true for a wide range of composite states, including those
bound together by the exchange of light vector or chiral multiplets.  The ground state then has
a non-degenerate radial wavefunction with $l=0$ \cite{Downs:1963} and factorizes as
\begin{eqnarray}
|\Psi\rangle =  |\psi(r)\rangle \otimes | \SS\rangle 
\label{eqn:StateProd}
\end{eqnarray}
to leading order, where $| \SS\rangle$ is the superspin part of the wavefunction.
At leading order in $v$ the supercharges act only on $|\SS\rangle$, leaving $|\psi(r)\rangle$ intact,
because gradients of non-relativistic wavefunctions are suppressed, $\partial_i \psi \sim \OO(v)$. 
Since $\psi(r)$ has trivial angular dependence, decomposing $| \SS\rangle $ into irreducible representations decomposes $|\Psi\rangle$ into irreducible supersymmetry representations,  $\Omega_j$, were $j$ refers to the spin of the Clifford vacuum (e.g.\;$\Omega_0$ is the chiral multiplet).
At $\OO(v^4)$ the Hamiltonian is typically spin-dependent and any degeneracy among the $\Omega_j$'s
will generically be lifted in the absence of any special symmetries.  For $\Omega_j$'s that are accidentally 
degenerate at $\OO(v^2)$, there can be large mixing that depends on the details of the dynamics,
though in many cases the appropriate mass eigenstates are determined by the action of addtional symmetries on
the supermultiplets.

As an example that illustrates this decomposition, consider the model bound state system that will form the main subject of this article.  It 
consists of four massive chiral superfields ($E$, $E^c$, $P$, $P^c$) with Dirac masses $m_e$ and $m_p$ satisfying
$m_e \le m_p$. The binding dynamics 
respect parity, under which the coordinates and superfields transform as
\begin{eqnarray}
x^\mu\leftrightarrow(-1)^\mu x^\mu\qquad \theta^\alpha\leftrightarrow\bar\theta_{\dot\alpha}\qquad P\leftrightarrow P^{c\dagger} \qquad E\leftrightarrow  
E^{c\dagger}
\end{eqnarray}
The dynamics also respect a $U(1)_R$-symmetry and a $U(1)_e \times U(1)_p$ flavor symmetry.
The charges of the component fields are taken to be
\begin{eqnarray}
\begin{array}{|c|cccccc|}
\hline
&\;\;\;\tilde p\;\;\;&\left(\begin{array}{c}p^\alpha\cr \bar p_{\dot\alpha}^c\end{array}\right)&\;\;\;\tilde p^{c\dagger}\;\;\;
&\;\;\;\tilde e\;\;\;&\left(\begin{array}{c}e^\alpha\cr \bar e^c_{\dot\alpha}\end{array}\right)&\;\;\;\tilde e^{c\dagger}\;\;\;
\cr
\hline
U(1)_R&1&0&-1&1&0&-1\cr
U(1)_{e+p}&1&1&1&1&1&1\\
\hline
\end{array}
\label{eqn:charges}
\end{eqnarray}
Significantly, the $U(1)_R$ symmetry and $\mathbb{Z}_{2}$ parity do {\em not} commute and combine into an $O(2)_R$ symmetry.
This can be seen by considering the ``selectrons" $\tilde{e}$ and $ \tilde{e}^{c}$. Parity,  $\PP$,  acts upon the selectrons as
\begin{eqnarray}
\mathcal{P} \tilde{e} = \tilde{e}^{c\dagger}
\end{eqnarray}
while under a $U(1)_R$ transformation, $R(\alpha)$, 
the selectrons transform as
\begin{eqnarray}
 R(\alpha) \tilde{e} = e^{i \alpha}\tilde{e} \qquad \text{and} \qquad R(\alpha) \tilde{e}^{c\dagger} = e^{-i \alpha}\tilde{e}^{c\dagger}
 \end{eqnarray}
so that $[\mathcal{P}, R(\alpha)] \tilde{e}  \ne 0$.
Thus $U(1)_R$ and $\mathbb{Z}_{2}$ are not a direct product and instead combine as the semi-direct product $U(1)_R \rtimes \mathbb{Z}_{2}  \cong O(2)_R$.  
This is important because $O(2)_R$ has two-dimensional irreducible representations that are realized in the bound state spectrum. In particular any state that transforms non-trivially under $U(1)_R$ must sit in an $O(2)_R$ doublet.

For this system, the superspin wavefunction $| \SS\rangle $ in Eq.~\ref{eqn:StateProd} decomposes as two chiral multiplets and one vector multiplet ($\VVV$),
as can be verified by counting degrees of freedom.   As a consequence of the $O(2)_R$ symmetry, however, the two chiral multiplets
combine into a hypermultiplet $\HHH$ so that the decomposition of $| \SS\rangle $ reads
\begin{eqnarray}
| \SS\rangle  = 2\Omega_0 \oplus \Omega_{\frac{1}{2}} = \HHH \oplus \VVV
\label{eqn:chiralxchiral}
\end{eqnarray}
Both $\HHH$ and $\VVV$ are charged under the global $U(1)_{e+p}$ flavor symmetry of the theory.
The superspin wavefunctions of the ground state are fixed by (super)symmetry at leading order
because $\HHH$ and $\VVV$ are irreducible under the full symmetry group and therefore insensitive to mixing.

Supersymmetry organizes non-relativistic pairs of \textit{free} particles  into supermultiplets,  determining the bound state wavefunctions at leading order in $v$ in terms of the constituent particles.
The organization of pairs of free particles  into supermultiplets is found by putting $E$, $E^c$, $P$ and $P^c$ on shell and constructing all possible superfield bilinears. 
The resulting bilinears will have spins ranging from 0 to 1. For example, the superfields $P$ and $E^{c\dagger}$ yield the bilinears   $PE^{c\dagger}$, $\DD^\alpha PE^{c\dagger}$, $P \bar \DD_{\dot\alpha}E^{c\dagger}$ and $\DD^\alpha P\bar \DD_{\dot \alpha}E^{c\dagger}$.  These bilinears can then be decomposed into irreducible supersymmetry representations 
with the help of projection operators, which in the case of spin zero superfields are given by 
\begin{eqnarray}
\PP_1=\frac{\DD^2\bar\DD^2}{16\Box} 
\qquad \PP_2=\frac{\bar\DD^2\DD^2}{16\Box} 
\qquad \text{and}\qquad
 \PP_T=-\frac{\DD\bar\DD^2\DD}{8\Box}
\label{eqn:projectionoperators}
\end{eqnarray}
where $\PP_1+\PP_2+\PP_T=1$ \cite{Wess:1992cp}. 

The decomposition is simplified by noting that the same state can appear in many different bilinears. In fact the bilinears
\begin{eqnarray}
\label{eqn:bilinears}
\PP_2PE=PE \qquad \PP_1 P^{c\dagger} E^{c\dagger}=P^{c\dagger} E^{c\dagger} \qquad\text{and}\qquad \PP_TPE^{c\dagger}
\end{eqnarray}
contain all the states as can be verified by counting degrees of freedom.  Expanding the first two bilinears in Eq.~\ref{eqn:bilinears} using the non-relativistic fields\footnote{The superscript $D$ indicates that the spinor is in the Dirac basis, where $\gamma_0$ is diagonal.}
\begin{eqnarray}
\tilde p=\frac{e^{im_p t}}{\sqrt{2m_p}}\phi_p\quad\text{ and }\quad 
\Psi_p^D=e^{im_pt}\left(\!\begin{array}{c}\psi_p\cr\frac{i\vec{\sigma}\cdot\vec{\nabla}}{2m_p}\psi_p\end{array}\!\right)
\label{eqn:non-relfields}
\end{eqnarray}
gives the superfields (cf. \cite{DiVecchia:1985xm})
\begin{eqnarray}
\label{eqn:pe}
PE\propto &&\phi_p\phi_e+\sqrt{2}\Theta^a\left(c_\theta\psi_p^a\phi_e+s_\theta\phi_p\psi_e^a\right)\\
\nonumber
&&-\Theta^2\left(s_\theta^2\phi_p\phi_{e^c}^\dagger+c_\theta^2\phi_{p^c}^\dagger\phi_e-s_{2\theta}(\psi_p\psi_e)_0\right)
\end{eqnarray}
and
\begin{eqnarray}
\label{eqn:pcec}
P^{c\dagger} E^{c\dagger}\propto &&\phi_{p^c}^\dagger\phi_{e^c}^\dagger+\sqrt{2}\bar\Theta_{a}\left(c_\theta\psi_{p}^{a}\phi_{e^c}^\dagger
+s_\theta\phi_{p^c}^\dagger\psi_{e}^{a}\right)\\
\nonumber
&&-\bar\Theta^2\left(c_\theta^2\phi_p\phi_{e^c}^\dagger+s_\theta^2\phi_{p^c}^\dagger\phi_{e}+s_{2\theta}(\psi_p\psi_e)_0\right)
\end{eqnarray}
where the dimensionless $\Theta^\alpha=\sqrt{m_p+m_e}\theta^\alpha$ has been introduced, 
and the mixing angle $\theta$ is defined by
\begin{eqnarray}
\tan^2 \theta = \frac{ m_e}{m_p} 
\end{eqnarray}
These two superfields have $U(1)_R$-charges of $\pm2$ and transform into each other under parity;
they correspond to the two $\Omega_0$'s in $\HHH$.
The $\VVV$ wavefunctions are found by decomposing the bilinear $\PP_T P E^{c\dagger}$, which gives a complex vector 
(curl) superfield with components
\begin{align}
D&\propto c_{2\theta}(\psi_p\psi_e)_0+s_{2\theta}(\phi_{p^c}^\dagger\phi_e-\phi_p\phi_{e^c}^\dagger)/{\sqrt{2}}\cr
\bar\lambda_1&\propto s_\theta\psi_p\phi_{e^c}^\dagger-c_\theta\phi_{p^c}^\dagger\psi_e\cr
\lambda_2&\propto s_\theta\psi_p\phi_e-c_\theta\phi_p\psi_e\cr
v^\mu&\propto\psi_p\vec{\sigma}\psi_e
\end{align}
Going to the parity eigenbasis and introducing notation for the various states gives
\begin{eqnarray}
 \nonumber
\VVV &=& \begin{cases}
v_\mu  \;\;\;\;\;\;\;\; & \ket{\vec{v}} = \ket {(\psi_p\psi_e)_1 }\ \ \\
\chi_+, \bar{\chi}^c_+&\ket{\psi_{\chi_+}}=c_\theta \ket{\phi_{p+} \psi_e} - s_\theta\ket{ \psi_{p} \phi_{e+}}\\
\chi_-, \bar{\chi}^c_- &\ket{\psi_{\chi_-}}= c_\theta\ket{\phi_{p-}\psi_e}-s_\theta\ket{\psi_p \phi_{e-} }\\
\varsigma_- &\ket{\varsigma_-}= c_{2\theta} \ket{(\psi_p \psi_e)_0} + \tfrac{s_{2\theta}}{\sqrt{2}} (\ket{\phi_{p+}\phi_{e-}} - \ket{\phi_{p-}\phi_{e+}})
\end{cases} \\
\ \HHH &=&\begin{cases}
\omega_+  \;\;\;\;\;\;\; & \ket{\omega_+} =  \tfrac{1}{\sqrt{2}}(\ket{\phi_{p+} \phi_{e+}} - \ket{\phi_{p-}\phi_{e-}}) \\
\omega_-&\ket{ \omega_-}= \tfrac{c_{2\theta}}{\sqrt{2}} (\ket{\phi_{p-}\phi_{e^+}} - \ket{\phi_{p+}\phi_{e-}})+s_{2\theta} \ket{(\psi_p \psi_e)_0 }\\
\xi_+, \bar{\xi}_+^c & \ket{\psi_{\xi_+}}=c_\theta \ket{\psi_p \phi_{e+}} + s_\theta \ket{\phi_{p+} \psi_e}\\
\xi_-, \bar{\xi}_-^c&\ket{\psi_{\xi_-}}= c_\theta \ket{\psi_p \phi_{e-}} + s_\theta \ket{\phi_{p-} \psi_e}\\
\varpi_+&\ket{ \varpi_+}=  \tfrac{1}{\sqrt{2}}(\ket{\phi_{p+} \phi_{e+}} +\ket{ \phi_{p-}\phi_{e-}})\\
\varpi_-& \ket{\varpi_-}=  \tfrac{1}{\sqrt{2}}(\ket{\phi_{p+} \phi_{e-}} + \ket{\phi_{p-}\phi_{e+}})
\end{cases}
\label{eqn:wavefunctions}
\end{eqnarray}
where $(\psi_p\psi_e)_0=\tfrac{1}{\sqrt{2}}\epsilon^{ab}\psi_p^a\psi_e^b$, $\phi_{p/e\pm}=\tfrac{1}{\sqrt{2}}(\phi_{p/e}\pm\phi_{p^c/e^c}^\dagger)$, and $c_\theta, s_\theta$  are  $\cos \theta$ and $\sin\theta$, respectively.  These are the same wavefunctions found in \cite{Rube:2009yc,Herzog:2009fw} by means of a detailed perturbative calculation in the particular case of supersymmetric hydrogen.

Although the $\VVV$ and $\HHH$ wavefunctions have been determined here without specifying the binding
dynamics, the mass splitting between $\VVV$ and $\HHH$ can only be determined by doing a dynamical
calculation.  In the absence of any special symmetries, however, it is expected that $m_{\text{FS}} \equiv m_{\VVV}-m_{\HHH}$ 
will be at the fine structure scale, $m_{\text{FS}} \sim \OO(v^4 \mu)$, and in the case of supersymmetric hydrogen one finds
\begin{eqnarray}
m_{\text{FS}} = \half \alpha_\vv^4 \mu.
\end{eqnarray}

The states in Eq.~\ref{eqn:wavefunctions} are organized according to their  $O(2)_R$ representations, with simple transformation properties under parity, because the breaking of $O(2)_R$ plays an important role in lifting degeneracies in the spectrum once supersymmetry is broken.
The states  $\chi_\pm$, $\xi_\pm$ and $\varpi_\pm$ transform in two-dimensional representations of $O(2)_R$ with $U(1)_R$ charges of 1, 1, and 2, respectively. 
For example, the doublet 
\begin{eqnarray}
\varpi = \begin{pmatrix}{\varpi_+}\\ {i \varpi_- } \end{pmatrix}
\end{eqnarray}
transforms irreducibly as
\begin{eqnarray}
\mathcal{P}:\varpi \to \sigma_3 \varpi  \qquad \qquad R(\alpha): \varpi \to e^{2 i \alpha \sigma_2}  \varpi
\end{eqnarray}
The states $v_\mu$, $\varsigma_-$ and $\omega_\pm$ are invariant under $R(\alpha)$ and thus transform as $O(2)_R$ singlets. 

To illustrate the action of supersymmetry on the ground states, consider  the heavy proton limit, $\theta \rightarrow 0$.  In this limit, supersymmetry clocks the states of the heavier constituent, leaving the valence particle intact.
 In particular, the $\VVV$ states consist of a light electron orbiting a heavy proton multiplet and the $\HHH$ states consist of a light selectron orbiting a heavy proton multiplet.  

This method of calculating superspin wavefunctions through decomposing products of superfields is general and can be applied to a 
wide class of non-relativistic supersymmetric bound state problems. For example, the superspin wavefunctions of non-relativistic $SU(3)$ baryons 
can be found by studying the decomposition of superfield trilinears. In this case, acting with the projection operators in Eq.~\ref{eqn:projectionoperators} 
on spin zero trilinears does not give all of the wavefunctions, and a spin $\half$ trilinear is necessary.   Similarly, the study of the bound states of a massive chiral and a massive vector superfield requires higher spin projections.

\subsubsection*{Excited state wavefunctions}
This prescription for finding the superspin wavefunctions can also be applied to the excited states. 
For a given spatial wavefunction $|nl\rangle$, the various excited states can be built by acting with supersymmetry
on the Clifford vacua defined by the particle content,
\begin{eqnarray}
\left\{|nl\rangle\otimes|\Omega_s\rangle,\quad
|nl\rangle\otimes\left(a^\dagger \otimes|\Omega_s\rangle\right),\quad
|nl\rangle\otimes\left(a^{\dagger2}|\Omega_s\rangle\right)\right\}
\label{eqn:ls-basis}
\end{eqnarray}
where $|\Omega_s\rangle$, $a^\dagger|\Omega_s\rangle$ and $a^{\dagger2}|\Omega_s\rangle$ are the superspin 
wavefunctions derived in the previous section.
For example in the case considered above $|\Omega_s\rangle$ is either $|\Omega_0\rangle$ or $|\Omega_{\half}\rangle$, and
the raising operators fill out the various states in $\VVV$ and $\HHH$.
Decomposing Eq.~\ref{eqn:ls-basis} into supermultiplets is equivalent to switching to the basis
\begin{eqnarray}
\left\{|nl\rangle\otimes|\Omega_s\rangle,\quad
a^\dagger\otimes\left(|nl\rangle\otimes|\Omega_s\rangle\right),\quad
a^{\dagger2}\left(|nl\rangle\otimes|\Omega_s\rangle\right)\right\},
\label{eqn:j-basis}
\end{eqnarray}
since irreducible representations of supersymmetry are obtained by acting with the raising
operator on Clifford vacua that are irreducible representations of the rotation group.  This
basis switch is just a matter of Clebsch-Gordon algebra and results in the decomposition
\begin{eqnarray} 
l\otimes\Omega_s=\Omega_{|l-s|}\oplus...\oplus\Omega_{|l+s|}.
\end{eqnarray}
For example, in the case considered above, where the bound state is formed from two chiral multiplets, 
the decomposition gives
\begin{eqnarray} 
\label{eqn:decompexample} 
l\otimes(\Omega_0\otimes\Omega_0)= l \otimes( \Omega_0 \oplus \Omega_0 \oplus \Omega_\half) =\Omega_{l-1/2}\oplus \Omega_l\oplus \Omega_l\oplus\Omega_{l+1/2}
\end{eqnarray} 
with the two $\Omega_l$
related to one another by parity.

Thus provided that a given $\Omega_j$ does not undergo large mixing, the excited state angular/superspin wavefunctions
can be found just as for the ground state.  One does a (single) superfield calculation as before to
determine $\Omega_s$ and then transforms from the basis of Eq.~\ref{eqn:ls-basis} to that in Eq.~\ref{eqn:j-basis} 
using Clebsch-Gordan coefficients.

\subsection{Effective Action for the Ground State}
\label{Sec: Effective Action}
Once the ground state spectrum is known, it is important to determine how the various
states interact with one another as well as with the SM.  
There are a variety of interactions, many of which are related through supersymmetric Ward identities. 
Superfields thus offer a convenient method for packaging all these interactions into manifestly supersymmetric forms. 
This section uses the standard off shell superfield formalism to formulate an effective free action for
the ground state, postponing until Sec.~\ref{Sec: Interactions} a discussion of ground state interactions.

$\HHH$ is described by  two chiral superfields that satisfy the following relations on shell 
\begin{eqnarray}
\HH_1\propto PE  \qquad \text{and} \qquad \HH_2^\dagger\propto P^{c\dagger}E^{c\dagger} .
\end{eqnarray}
A second set of chiral superfields, $\HH_1^{c\dagger}$ and $\HH_2^c$, is introduced
to give the $F$-terms of $\HH_1$ and $\HH_2^\dagger$ dynamics.
The free Lagrangian for $\HHH$ is given by
\begin{eqnarray}
\label{eqn:haction}
\LL_{\HHH} = \int\!\!d^4\theta\;  \delta^{ij}\!\! \left( \HH^\dagger_i \HH_j +  \HH^c{}^\dagger_i \HH_j^c\right)
+\int\!\! d^2\theta \;  \delta^{ij} m_{\HH} \HH_i \HH^c_j +\!\!\hc 
\label{eqn:Haction}
\end{eqnarray}
The equations of motion which follow from Eq.~\ref{eqn:haction} then result in the identification
\begin{eqnarray}
\HH_1^{c\dagger} \propto \mathcal{P}_1 P^{c\dagger} E \qquad \text{and} \qquad \HH_2^c\propto \mathcal{P}_2 P^{c\dagger} E
\end{eqnarray}
With $\HH_i$ and $\HH_i^c$ identified as above the appropriate $U(1)_R$ and $U(1)_{e+p}$ charges are given by
\begin{eqnarray}
\begin{array}{|c|cccc|}
\hline
& \HH_1 & \HH_2^\dagger & \HH_1^{c\dagger} & \HH_2^c \\
\hline
U(1)_R& 2 & -2 & 0 & 0 \\
U(1)_{e+p}& 2 & 2 & 2 & 2 \\
\hline
\end{array}
\end{eqnarray}
so that the Lagrangian is properly invariant under $U(1)_R$ and $U(1)_{e+p}$.
Parity acts on the composite superfields as
\begin{eqnarray}
\HH_1\leftrightarrow \HH_2^{\dagger}\;\;\qquad\;\;\HH_1^c\leftrightarrow \HH_2^{c\dagger} .
\end{eqnarray}
so that the Lagrangian is also invariant under parity.

$\VVV$ is described by an off shell field, $\VV$, and an action consistent with the on shell constraint $\VV\propto \PP_T PE^{c\dagger}$. $\VV$ is a charged vector superfield -- a general superfield with no Lorentz index 
\begin{eqnarray}
\VV(x, \theta,\bar{\theta}) \ne \VV^\dagger(x,\theta,\bar{\theta}).
\end{eqnarray}
 The action is written with the help of the supersymmetric field strengths
\begin{eqnarray}
\WW_{1\,\alpha} =  -\frac{1}{4}\, \bar{\DD}^2 \DD_\alpha \VV \qquad \text{and} \qquad \WW_{2\,\alpha} = -\frac{1}{4}\,\bar{\DD}^2 \DD_\alpha \VV^\dagger 
\end{eqnarray}
which have $U(1)_R$ charges of $+1$.  Under parity $\VV$ and $\WW_i$ transform as 
\begin{eqnarray}
\VV\leftrightarrow -\VV  \qquad\text{and}\qquad \WW_1^\alpha \leftrightarrow -\bar{\WW}_{2\dot\alpha} 
\end{eqnarray}
The free Lagrangian, which is properly invariant under $U(1)_R$ and parity, is given by
\begin{eqnarray}
\LL_{\VVV} = \int\!\!d^4\theta\;  2\,m_{\VV}^2 \VV^\dagger  \VV + \int\!\!d^2\theta\; \frac{1}{2} \WW_1^\alpha \WW_{2\,\alpha} +\!\hc 
\end{eqnarray}
Varying the action yields the equation of motion $\DD_\alpha\WW_1^\alpha=2\,m_\vv^2\VV$, implying that $\PP_T \VV=\VV$ on shell. 

\section{Supersymmetry Breaking in the Ground State}
\label{Sec: SSB Effects}

The previous section calculated the composition of non-relativistic supersymmetric bound states using supersymmetric group theory,
focusing on the particular example of bound states formed from two chiral superfields. 
 This section builds on Sec.~\ref{Sec: NR} by incorporating the effects of weak supersymmetry breaking on the ground state spectrum. 
The exact changes to the spectrum resulting from supersymmetry breaking depend on the details of the binding dynamics. 
In many theories, however, supersymmetry breaking level splittings induced by the binding dynamics are accompanied by powers of the velocity, $v$.  Consequently in the non-relativistic limit supersymmetry breaking in the bound state spectrum 
will be dominated by the differences in the rest energies of the constituent fermions and bosons.
For such theories the resulting spectrum is insensitive to the details of the binding dynamics.

\subsection{Constituent Mass Effects}
\label{Sec: CME}

The leading supersymmetry breaking effects can be calculated by folding in the perturbed rest energies of the constituents
with the ground state superspin wavefunctions calculated in Sec.~\ref{Sec: NR}.
 This leading order effect is straightforward to compute if the effective scale of supersymmetry breaking in the bound states spectrum, $m_{\text{soft}}$, is smaller than the scale of principle excitations
\begin{eqnarray}
m_{\text{soft}} \ll m_{\text{prin}} \simeq \OO(\mu v^2) .
\end{eqnarray}
In this case mixing with excited states is unimportant and the incorporation of supersymmetry breaking into the bound state spectrum reduces to a finite dimensional quantum mechanical perturbation theory problem.

The bound state spectrum has two effective mass scales for supersymmetry breaking effects.
The first scale is set by the $U(1)_R$-preserving soft masses, $m_{R-{\text{pres}}}$, 
while  the second is set by the $U(1)_R$-violating $B$-term masses, $m_{R-{\text{viol}}}$.  The breaking of
the $U(1)_R$ symmetry induces splittings between states that are doublets under the $O(2)_R$ symmetry.
In many implementations of dark sector supersymmetry breaking, $U(1)_R$-violating soft terms will be suppressed 
relative to the $U(1)_R$-preserving soft terms and for simplicity the relative ordering of the scales is taken to be
  \begin{eqnarray}
  \label{assumption}
m_{R-\text{pres}}, m_{\text{FS}}\gg m_{R-\text{viol}} 
 \end{eqnarray}
 throughout, where $m_{\text{FS}} = \OO(\mu v^4)$.  
 
The soft supersymmetry breaking Lagrangian for the chiral-chiral bound state system introduced in Sec.~\ref{Sec: NR} 
contains a $U(1)_R$-preserving piece,
\begin{eqnarray}
\label{eqn:lpreserve}
-\LL_{R-\text{pres}} \supset \Delta^2_{\tilde{e}} (|\tilde{e}|^2 + |\tilde{e}^c|^2) + \Delta^2_{\tilde{p}} (|\tilde{p}|^2 + |\tilde{p}^c|^2)
\end{eqnarray}
and additional supersymmetry breaking terms that break the $U(1)_R$ symmetry:
\begin{eqnarray}
\label{eqn:lbreak}
-\LL_{R-\text{viol}}  \supset  B_e m_e \tilde{e} \tilde{e}^c + B_p m_p \tilde{p}\tilde{p}^c +\!\!\hc 
\end{eqnarray}
For simplicity the soft parameters are assumed to obey the relations
\begin{eqnarray}
\Delta^2_{\tilde{e}} \simeq \Delta^2_{\tilde{p}} \qquad\text{ and }\qquad B_p\simeq B_e \equiv B
\end{eqnarray}
In the presence of $\Delta^2_{\tilde{e}}$ and $B_e$ the selectron mass eigenstates become 
\begin{eqnarray}
\tilde{e}_\pm = \frac{1}{\sqrt 2}(\tilde{e}\pm\tilde{e}^{c\dagger})
\end{eqnarray}
with masses
\begin{eqnarray}
m_{\tilde{e}\pm} = m_e + \delta m_{\tilde{e}_\pm}  \equiv m_e + \half \frac{ \Delta^2_{\tilde{e}}}{m_{\text{e}}}  \pm \half B 
\end{eqnarray}
Analogous expressions hold for the mass eigenstates $\tilde{p}_\pm$. 
See Sec.~\ref{Sec: SSB} for details on a particular implementation of supersymmetry breaking in the dark sector that
satisfies the above assumptions.

The leading supersymmetry breaking perturbation on the ground state spectrum is encapsulated in
the perturbing Hamiltonian
\begin{eqnarray}
H_{\text{soft}} =   \delta m_{\tilde{p}\pm} |\phi_{p\pm}\rangle \langle\phi_{p\pm} | + \delta m_{\tilde{e}\pm } |\phi_{e\pm}\rangle\langle\phi_{e\pm} |
\end{eqnarray}
The $U(1)_R$-preserving contributions of $H_{\text{soft}}$ will appear in the combination
\begin{eqnarray}
m_{\text{soft}} \equiv \half (\delta m_{\tilde{p}+}+\delta m_{\tilde{p}-} + \delta m_{\tilde{e}+}+\delta m_{\tilde{e}-})
\end{eqnarray}
The rest energy perturbations can now be read off directly from the supersymmetric wavefunctions in Eq.~\ref{eqn:wavefunctions}.
For example consider the state
\begin{eqnarray}
\ket{\varsigma_-}= c_{2\theta} \ket{(\psi_p \psi_e)_0} + s_{2\theta} (\ket{\phi_{p+}\phi_{e-}} - \ket{\phi_{p-}\phi_{e+}})/\sqrt{2}
\end{eqnarray}
The fermion-fermion component is insensitive to $H_{\text{soft}}$, but the scalar-scalar component results in a perturbation
\begin{eqnarray}
\Delta m_{\varsigma_-} \simeq \langle \varsigma_- | H_{\text{soft}} |\varsigma_-\rangle = 
\half s_{2\theta}^2 (\delta m_{\tilde{p}+}+\delta m_{\tilde{e}-} + \delta m_{\tilde{p}-}+\delta m_{\tilde{e}+}) = s^2_{2 \theta} \; m_{\text{soft}}
\end{eqnarray}
 which is the leading supersymmetry breaking contribution to the mass of $\varsigma_-$ in the limit that
 $m_{\text{FS}} \gg m_{\text{soft}}$. For many physical applications, such as decays or scattering off of SM nuclei, knowing only the leading breaking is sufficient. Using the superspace approach for finding the wavefunctions, as in Sec.~\ref{Sec: Wavefunctions Fields}, the leading supersymmetry breaking can thus be found for a broad range of perturbative bound states.

\subsection{Subdominant Effects}
\label{Sec: Subdominant}

Supersymmetry breaking effects begin to grow in complexity beyond the rest mass perturbation.
The next most important term in the non-relativistic expansion is the kinetic energy perturbation
\begin{eqnarray}
H_{v^2}  \simeq -\frac{p^2}{2\mu} \frac{\delta \mu}{\mu }
\end{eqnarray}
This changes the principle structure of the bound state and leads to a $\OO(v^2)$ perturbing Hamiltonian 
\begin{eqnarray}
\label{Eq: KineticPert}
H_{v^2} = -\left \langle \frac{p^2}{2\mu^2} \right\rangle
\left( \cos^4\!\theta \; \delta m_{\tilde{e}\pm} |\phi_{e\pm}\rangle\langle \phi_{e\pm}| + \sin^4\! \theta \; \delta m_{\tilde{p}\pm}  |\phi_{p\pm}\rangle\langle \phi_{p\pm}|
\right) .
\end{eqnarray}  
At the level of fine structure many new effects arise.  These include additional kinematic effects from $\OO(p^4)$ terms and, in
the case of supersymmetric hydrogen, gaugino mass effects and $D$-term contributions.
Incorporating all these effects requires using the $\TT$ matrix and computing all tree-level Feynman diagrams contributing to $ep \rightarrow ep$ matrix elements.
The $\TT$ matrix is proportional to an effective non-relativistic Hamiltonian that can be used to do perturbation theory,
as in the calculation of the fine structure of supersymmetric hydrogen \cite{Buchmuller:1981bp, Rube:2009yc,Herzog:2009fw}.

\subsection{Eigenstates} 

In this section the ground state spectrum with weakly broken supersymmetry is presented by diagonalizing
the perturbation $H_{\text{soft}}$.

 \subsubsection*{Scalars}
 
 In the absence of supersymmetry breaking the hypermultiplet contains the degenerate pair of positive parity scalar bound states $\varpi_+$ 
 and $\omega_+$.  In the presence of $H_{\text{soft}}$ these states mix maximally:
\begin{eqnarray}
\left(\begin{array}{ccc}
\varpi_+  \\ \omega_+   \end{array}\right)^{\!\!\! \dagger}
\left(\begin{array}{ccc}
m_{\text{soft}}  &   B \\
B   &  \;\;\;   m_{\text{soft}} 
\end{array}\right)
\left(\begin{array}{ccc}
\varpi_+  \\ \omega_+   \end{array}\right)
\end{eqnarray}
$B$ characterizes the size of $O(2)_R$ breaking and mixes states of different $R$-charge.
The mass eigenstates are 
\begin{eqnarray}
\nonumber
\label{eqn:omegaoneplus}
\omega_{1+} &\equiv& \phi_{p+}\phi_{e+}  = \frac{1}{\sqrt 2}(\varpi_+ + \omega_+)
\qquad\quad m_{\omega_{1+}} =  \delta m_{\tilde{e}+}+\delta m_{\tilde{p}+}=m_{\text{soft}} +  B \\
\omega_{2+} &\equiv& \phi_{p-}\phi_{e-}   = \frac{1}{\sqrt 2}(\varpi_+ - \omega_+)
\qquad\quad m_{\omega_{2+}} = \delta m_{\tilde{e}-}+\delta m_{\tilde{p}-}= m_{\text{soft}} -B  .
\end{eqnarray}

\FIGURE{
\centering
\qquad\includegraphics[width=3.20in]{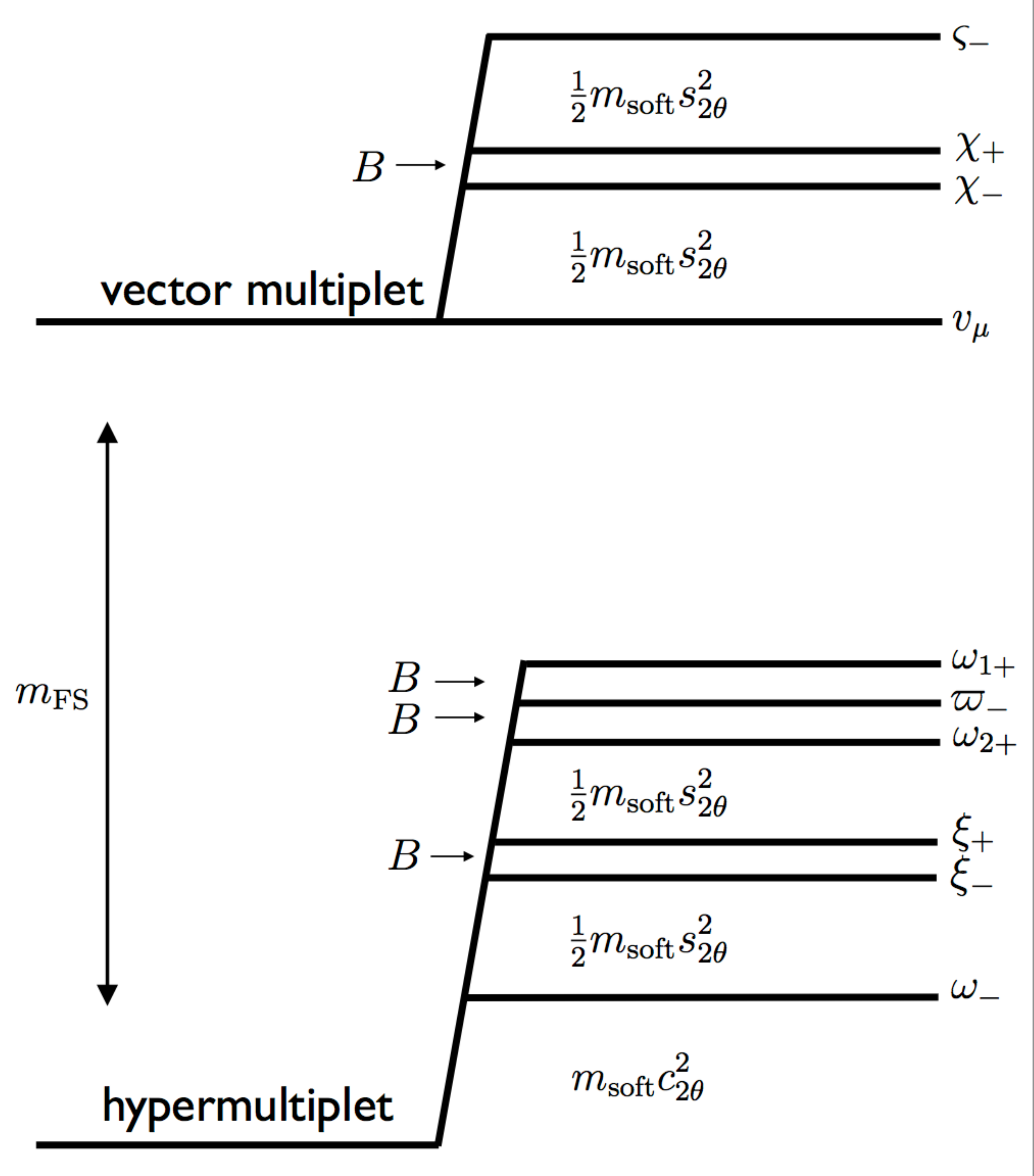}\qquad
\caption{The ground state spectrum for the case $m_{\text{FS}} > 0$ and in the limit that $m_{\text{FS}} \gg m_{\text{soft}}$, where the mixing between the vector 
multiplet and hypermultiplet is small. The composition of the various states (see Eq.~\ref{eqn:wavefunctions} 
and Eq.~\ref{eqn:omegaoneplus}) is a function of $\tan^2  \theta = m_e/m_p$.  For example, in the heavy proton limit 
($m_e \ll m_p$), $\omega_-$ is predominantly scalar-scalar, whereas in the opposite limit ($m_e \simeq m_p$), 
$\omega_-$ is predominantly fermion-fermion.}}

In the supersymmetric limit the ground state contains three parity odd scalars, one of which,
 $\varsigma_-$, is in the vector multiplet and two of which, $\omega_-$ and $\varpi_-$, are in
 the hypermultiplet.  In the presence of $H_{\text{soft}}$ and $H_{v^2}$ all
 three states mix:
 \begin{eqnarray}
  \label{3x3matrix}
\left(\begin{array}{ccc}
\varsigma_- \! \\  \omega_-\! \\ \varpi_-\!  \end{array}\right)^{\!\!\! \dagger}\!\!
  \left( \begin{array}{ccc}
 m_{\text{FS}} + s_{2\theta}^2  m_{\text{soft}}  & -\half s_{4\theta}  m_{\text{soft}} &\half s_{4\theta}  B^{\prime} \\
-\half s_{4\theta} m_{\text{soft}} & c_{2\theta}^2  m_{\text{soft}}  &-c_{2\theta}^2 B^{\prime} \\
\half s_{4\theta}  B^{\prime} & -c_{2\theta}^2 B^{\prime} &m_{\text{soft}}   
\end{array} \right)
  \left(\begin{array}{ccc}
\varsigma_- \! \\  \omega_- \! \\ \varpi_-\!  \end{array}\right)
 \end{eqnarray}
Here  $B^{\prime}$ characterizes the $U(1)_R$-breaking in this sector and comes about through $H_{v^2}$ in Eq.~\ref{Eq: KineticPert} or from the difference in the $B$-term masses between the $\tilde{e}$ and $\tilde{p}$ and is of the order 
\begin{eqnarray}
\label{negparity1}
B' \sim \OO(Bv^2), \OO(B_e - B_p).
\end{eqnarray}
By specializing to the regime where 
\begin{eqnarray}
B^{\prime} \ll m_{\text{FS}}, \;\tan^2 2\theta \,m_{\text{soft}} \qquad
\end{eqnarray}
so that mixing between the three states occurs primarily between $\varsigma_-$ and $\omega_-$, one obtains simple formulae for the approximate energy levels:
\begin{eqnarray}
\label{eqn:varsigma-}
m_{\varsigma_-}& =&   
\frac{m_{\text{FS}} + m_{\text{soft}}}{2} + \left (\frac{m_{\text{FS}} - c_{4\theta}m_{\text{soft}}}{2} \right)  
\sqrt{1+\frac{ s_{4\theta}^2 \; m_{\text{soft}}^2 }{ (m_{\text{FS}} - c_{4\theta}m_{\text{soft}})^2}}\\
\label{eqn:omega-}
m_{\omega_-}& =&\frac{m_{\text{FS}} + m_{\text{soft}}}{2} - \left (\frac{m_{\text{FS}} - c_{4\theta} m_{\text{soft}}}{2} \right)  
\sqrt{1+\frac{ s_{4\theta}^2 \;  m_{\text{soft}}^2}{ ( m_{\text{FS}} - c_{4\theta}m_{\text{soft}})^2}}  \\
m_{ \varpi_-}& =& m_{\text{soft}}  
\end{eqnarray}
Non-zero $B$-terms split the $O(2)_R$ doublet containing $ \varpi_+$ and $ \varpi_-$.
   
In the limit $m_{\text{soft}}/m_{\text{FS}}\rightarrow \infty$ with $0 \le \theta < \tfrac{\pi}{8}$ (respectively $\tfrac{\pi}{8} < \theta \le \tfrac{\pi}{4}$), 
the state $\varsigma_-$ (respectively $\omega_-$) becomes the $(\psi_e \psi_p)_0$ bound state.  Naively, for $\theta^2 \simeq \frac{m_e}{m_p} \simeq \tfrac{1}{1836}$ and $m_{\text{FS}}=\half \alpha_\vv^4 \mu$
the splitting between this state and the vector $(\psi_e \psi_p)_1$ should give the hyperfine splitting in regular hydrogen; 
however, Eq.~\ref{eqn:varsigma-} yields instead:
\begin{eqnarray}
\label{Eq: HyperfineEmergence}
m_{(\psi_e \psi_p)_1} - m_{(\psi_e \psi_p)_0} = m_{v_\mu} - m_{\varsigma_-} \rightarrow s_{2\theta}^2 m_{\text{FS}} \rightarrow \frac{2\alpha_\vv^4  m_e^2}{ m_p} 
\end{eqnarray}   
This is not the correct  hyperfine splitting of regular hydrogen which is
\begin{eqnarray}
m_{\text{HFS}} = \frac{8}{3} \frac{ \alpha^4 m_e^2}{m_p}
\end{eqnarray}
for a point-like proton.
  This difference arises because
$m_{\text{soft}}/m_{\text{FS}}\rightarrow \infty$ is {\it not} the full decoupling limit.   
In particular, the ground state of supersymmetric hydrogen contains admixtures of higher principle excitations arising from 
gaugino exchange at second order in perturbation theory.   
These effects contribute to the hyperfine splitting in Eq.~\ref{Eq: HyperfineEmergence} but disappear in the full decoupling limit where the gaugino mass goes to infinity, $m_{\tilde{\vv}}\rightarrow \infty$. 
   
\subsubsection*{Fermions}
In the absence of supersymmetry breaking the vector multiplet (hypermultiplet) contains the degenerate
pair of $j=\half$ bound states $\xi_\pm$ ($\chi_\pm$).   In the presence of $H_{\text{soft}}$ the states of equal
parity mix with one another:
\begin{eqnarray}
 \left(\begin{array}{ccc}
\bar{\chi}_\pm \\  \bar{\xi}_\pm  \end{array}\right)^{\!\!\! T}
\left(\begin{array}{ccc}
m_{\text{FS}} +\half s_{2\theta}^2 m_{\text{soft}}  \pm\half B  & -\tfrac{1}{4} s_{4\theta} m_{\text{soft}}  \\
-\tfrac{1}{4} s_{4\theta} m_{\text{soft}}  & (s_{\theta}^4+c_{\theta}^4) m_{\text{soft}}  \pm\half B 
\end{array}\right)
 \left(\begin{array}{ccc}
\chi_\pm \\  \xi_\pm  \end{array}\right)
\label{eqn:fermionmatrix}
\end{eqnarray}
The spectrum is given by 
\begin{eqnarray}
m_{\chi_\pm}& =&   
\frac{m_{\text{FS}} + m_{\text{soft}}}{2} + \left (\frac{m_{\text{FS}} - c_{2\theta}^2 m_{\text{soft}}}{2} \right)  
\sqrt{1+\frac{\tfrac{1}{4} s_{4\theta}^2  m_{\text{soft}}^2 }{ (m_{\text{FS}} - c_{2\theta}^2 m_{\text{soft}})^2}}\pm \half B \\
m_{\xi_\pm}& =&\frac{m_{\text{FS}} + m_{\text{soft}}}{2} - \left (\frac{m_{\text{FS}} - c_{2\theta}^2 m_{\text{soft}}}{2} \right)  
\sqrt{1+\frac{\tfrac{1}{4} s_{4\theta}^2  m_{\text{soft}}^2 }{ (m_{\text{FS}} - c_{2\theta}^2 m_{\text{soft}})^2}}\pm \half B 
\end{eqnarray}
For non-zero $B$ the $O(2)_R$ symmetry that ensured the degeneracy of the pair of states $\chi_{\pm}$ as well as the pair of states $\xi_{\pm}$ is broken and the fermionic spectrum splits completely.  

\subsubsection*{Vector}
The vector state $\ket{\vec{v}} = \ket {(\psi_p\psi_e)_1 }$ is insensitive to $H_{\text{soft}}$ and, as a consequence,
does not feel supersymmetry breaking at leading order.

\section{Interactions}
\label{Sec: Interactions}

Composite systems have a wide range of interactions that are controlled by selection rules and form factors that results in these systems having a much richer phenomenology than elementary particles.     
This section uses the effective field theory of Sec.~\ref{Sec: Effective Action} to study the interactions 
that arise when composite states inherit gauge interactions from their constituents (cf. \cite{Bagnasco}).

Sec.~\ref{Sec: U1v} considers the case  where the constituents are charged under an unbroken vectorial gauge symmetry $U(1)_\vv$ such that the composite state is neutral with the following charge and parity assignments 
\begin{eqnarray}
\begin{array}{|c|cccc|}
\hline
& E& E^c& P & P^c\\
\hline
U(1)_\vv& -1& +1 & + 1& -1\\
\hline
\end{array} 
\qquad
\text{ and }\qquad \vv \leftrightarrow -\vv
\end{eqnarray}
$\vv$ does not need to be responsible for binding the chiral multiplets together; {\em e.g.}, the binding could arise from a Yukawa force.
 The $U(1)_\vv$ gauge interactions of the constituents induce a number of effective operators, including charge radius, Rayleigh scattering,  and spin flip operators.  Specializing to the case where the hypermultiplet $\HH$ is lighter than the vector multiplet $\VV$ and $m_{\text{FS}} \gg m_{\text{soft}}$, decays within the ground state are discussed in detail.  It is found that the states of  the vector multiplet $\VV$ decay relatively quickly down to $\HH$, while the decays within $\HH$ are much slower.  

Sec.~\ref{Sec: U1a} briefly considers
the case where the constituents are charged under a broken axial gauge symmetry $U(1)_\aa$ with charge and parity assignments
\begin{eqnarray}
\label{Eq: Axial Charges}
\begin{array}{|c|cccc|}
\hline
& E& E^c& P & P^c\\
\hline
U(1)_\aa& +1 & +1 & -1 & -1\\
\hline
\end{array} \qquad\text{ and }\qquad \aa \leftrightarrow \aa
\end{eqnarray} 
In models such as that of Sec.~\ref{Sec: Nearly Susy} where $\aa$ undergoes kinetic mixing with the SSM,
these interactions mediate the dominant coupling of dark atoms to the Standard Model. 
Sec.~\ref{Sec: U1a} discusses the allowed scattering channels and finds the leading supersymmetric axial interactions.

\subsection{$U(1)_\vv$ interactions}
\label{Sec: U1v}

The interactions of neutral bound states with an external vector superfield, $\vv$,
are characterized by two scales
corresponding to the charge radius, $R_e$, and magnetic radius, $R_m$.  Physically $R_e$ corresponds 
to the size of the bound state, $R_e \sim \sqrt{\langle r^2 \rangle}$.  In the case of Coulombic bound states $R_e$ is given by the Bohr radius, $R_e^{-1} =  \alpha_\vv \mu$.   $R_m$ is just the Compton wavelength, $R_m^{-1} = \mu$.  
For convenience, this section will restrict its discussion to   supersymmetric
hydrogen, although it is generally applicable to chiral-chiral bound states.

Before considering the supersymmetric case, it is instructive to review the leading interactions of the photon with the 
spin-singlet ground state of regular hydrogen. The leading elastic interaction comes from the charge radius operator
\begin{align}
g_\vv c_{2\theta} R_e^2(\psi_p\psi_e)_0^\dagger\partial_\mu(\psi_p\psi_e)_0 \partial_\nu \vv^{\mu\nu}
\label{eqn:charge-radius-ff}
\end{align}
which is fully determined by the charge distribution of the bound state. The leading inelastic interaction comes from the magnetic spin-flip operator which is determined by the fermion content:  
\begin{eqnarray}
g_\vv R_m\partial_\mu(\psi_p\psi_e)_{1,\nu}(\psi_p\psi_e)_0\tilde \vv^{\mu\nu}
\label{eqn:magnetic1}
\end{eqnarray}
Finally there is the Rayleigh scattering operator
\begin{eqnarray}
g_\vv R_e^3m_H(\psi_p\psi_e)_0^\dagger(\psi_p\psi_e)_0 \vv_{\mu\nu}\vv^{\mu\nu}
\label{eqn:rayleigh}
\end{eqnarray}
which makes the sky blue.  All other operators are higher order in either $g_\vv$ or $\mu^{-1}$. 

The next step is to find the set of operators necessary to satisfy the supersymmetric Ward identities.
 $\VV\leftrightarrow \VV$ interactions can be important for scattering processes
if the states of $\VV$ are long-lived but have a subdominant effect on the lifetimes of the states
in $\VV$.   Because the leading $\VV \leftrightarrow \HH$ decay is relatively fast $\VV$ tends to be short-lived 
and therefore $\VV\leftrightarrow \VV$ interactions are ignored here.   The $\HH \leftrightarrow \HH$ interactions are
found in the following, since they determine the relaxation timescale of the $\HH$ supermultiplet.     

The charge radius operator in Eq.~\ref{eqn:charge-radius-ff} only depends on the charge distribution, 
and therefore the scalar-scalar bound states must share identical (diagonal) interactions:
\begin{eqnarray}
g_\vv c_{2\theta} R_e^2(\phi_{p\pm}\phi_{e\pm})^\dagger \partial_{\mu} (\phi_{p\pm}\phi_{e\pm}) \partial_\nu \vv^{\mu\nu}
\label{eqn:charge-radius-ss}
\end{eqnarray}
Rewriting  Eq.~\ref{eqn:charge-radius-ff} and \ref{eqn:charge-radius-ss} in terms of the wavefunctions in Eq.~\ref{eqn:wavefunctions}, the charge radius interactions become
\begin{eqnarray}
g_\vv c_{2\theta} R_e^2\left(\omega^\dagger_\pm\partial_\mu\omega_\pm+\varpi^\dagger_\pm\partial_\mu\varpi_\pm+
\varsigma^\dagger_-\partial_\mu\varsigma_-\right)\! \partial_\nu \vv^{\mu\nu}
\label{eqn:charge-radius-SUSY1}
\end{eqnarray}
Similarly, the spin-flip and Rayleigh scattering operators become
\begin{eqnarray}
g_\vv R_m\partial_\mu v_\nu (c_{2\theta}\varsigma_-+s_{2\theta}\omega_-)\tilde \vv^{\mu\nu}
\label{eqn:spin-flip-SUSY1}
\end{eqnarray}
and 
\begin{eqnarray}
g_\vv R_e^3(m_e+m_p)\left(\omega^\dagger_\pm\omega_\pm+\varpi^\dagger_\pm\varpi_\pm+
\varsigma^\dagger_-\varsigma_-\right)\!\vv_{\mu\nu}\vv^{\mu\nu},
\label{eqn:rayleigh-SUSY}
\end{eqnarray}
respectively.
The operators in Eq.~\ref{eqn:charge-radius-SUSY1} to \ref{eqn:rayleigh-SUSY} represent the leading single photon
and two-photon interactions for the scalar states in $\VV$ and $\HH$.  
Several interactions remain to be found, {\em e.g.}, the leading single photino interactions as well as the interactions for the fermionic states. 
The coefficients of the remaining interactions are found by forming operators from the effective fields of Sec.~\ref{Sec: Effective Action}.
The matching coefficients are determined by expanding the supersymmetric operators in terms of their components and identifying the corresponding interactions from  Eq.~\ref{eqn:charge-radius-SUSY1} to \ref{eqn:rayleigh-SUSY}. 
This procedure allows for the various supersymmetric interactions to be systematically enumerated by building upon the known interactions of regular hydrogen.

\subsubsection*{Interactions of the Hypermultiplet with Higher States}

A variety of processes cause the decay of the excited states to the ground state.  For example, supersymmetric hydrogen inherits the (fast) electric dipole and magnetic dipole transitions
of regular hydrogen.  Decays from $\VV$ to $\HH$, however, are not as fast and merit further discussion.

The states in $\VV$ are connected to $\HH$ through two one-photon operators of dimension five:
\begin{eqnarray}
\label{Eq: HV Int1}
\LL_{\VV \HH \vv} &=&c_M s_{2\theta}\,g_\vv R_m\!  \int\!\! d^2\theta\; (\HH_1^c\WW_1+\HH_2^c\WW_2)\WW_{\vv}+\hc\\
\label{Eq: HV Int2}
&&+ c_{M}^{\prime}(g_\vv, \theta) g_\vv R_m\! \int\!\! d^4\theta\; (\HH_1^c+\bar\HH_2^c)\VV \DD\WW_\vv+\hc 
\end{eqnarray}
These two operators are the most general forms for $\HH\leftrightarrow\VV$ interactions mediated by  $U(1)_\vv$.  Higher dimensional operators can be reduced to these two forms with additional factors of $\partial^2$ acting on $\WW_\vv$ by using the matter field equations of motion. 

Only Eq.~\ref{Eq: HV Int1} contains the magnetic spin-flip interaction, $\omega_-\partial_\mu v_\nu \tilde F^{\mu\nu}_\vv$.
The factor of $s_{2\theta}$ is fixed by comparison with  Eq.~\ref{eqn:magnetic1}.
In supersymmetric hydrogen, some of the component interactions contained
in Eq.~\ref{Eq: HV Int1} arise from $\OO(\alpha_\vv)$ mixing between the ground state and higher principle excitations. 
For example, excited $\omega_+$ states ($2p$, $3p$, etc.)\! mix with $v^\mu$, allowing for $v^\mu$ to decay to $\omega_+$ through 
electric dipole transitions.  This mixing with excited states is the origin of the ``electric'' interaction $\omega_+\partial_\mu v_\nu F^{\mu\nu}_\vv$ contained in the operator of Eq.~\ref{Eq: HV Int1}.  In this sense, the operator of Eq.~\ref{Eq: HV Int1} is neither purely magnetic or electric.
The role that excited state mixing plays in ensuring this supersymmetric result is familiar from the calculation
of the supersymmetric spectrum in \cite{Buchmuller:1981bp,Rube:2009yc,Herzog:2009fw}, where second order perturbation theory  is needed to determine the spectrum to $\OO(\alpha_\vv^4)$.

The operator in Eq.~\ref{Eq: HV Int2} does not mediate decays in the supersymmetric limit.  This can be seen by 
using the equations of motion to replace 
$\DD\WW_\vv$ with the current $\JJ_\vv$. 
Decays through this operator are kinematically forbidden 
because the mass splitting between $\VV$ and $\HH$ is much smaller than the mass of any particle  charged under $U(1)_\vv$.
For this reason we leave the coefficient of this operator
undetermined, noting however that it can only come in at higher order than $g_\vv R_e$, since it contains off-diagonal scalar-scalar transitions, which
do not arise from charge radius scattering.

The various decay channels induced by the interactions in Eq.~\ref{Eq: HV Int1} cause each state in $\VV$ to have the same inclusive decay  width to the states of $\HH$ in the supersymmetric limit---otherwise the component propagators of $\VV$ would have different poles. Therefore, the decay width can be calculated by considering the state with the simplest decay modes, in this case $\varsigma_-$:
\begin{eqnarray}
\LL_{\VV \HH \vv} 
\supset 
c_M g_\vv R_m m_\vv \; \frac{i s_{2\theta} }{2\sqrt{2}}\left(i\bar\xi_+\gamma_5+\bar\xi_-\right) \Lambda_\vv\,\varsigma_-
\end{eqnarray}
Here $\Lambda_\vv$  is the four-component Majorana gaugino of $U(1)_\vv$ and $m_{\tilde{\vv}}$ is its mass.
This gives the decay rate 
\begin{eqnarray}
\Gamma_{\VV\rightarrow\HH \vv} \simeq |c_M g_\vv s_{2\theta} R_m m_\vv|^2  \frac{ m_{\text{FS}}^2}{4\pi m_\vv } 
   = |c_M|^2 \alpha_\vv^9 \mu\; 
\label{eqn:spinflipdecayrate}
 \end{eqnarray}
This is a factor of $\frac{\mu}{m_{\text{FS}}}$ faster than the corresponding spin-flip transition in regular hydrogen, which scales as $\alpha  m_{\text{FS}}^3$.    This is because the decays are dominated by $\Lambda_\vv$ emission rather than $\vv^\mu$ emission, for which the amplitude carries an additional factor  factor of $E^\half$, where $E$ is the energy of the emitted gauge particle. 

\subsubsection*{Interactions within the Hypermultiplet}

Supersymmetry restricts the form of possible interactions significantly, and  these restrictions are particularly severe for interactions connecting two chiral superfields.  For instance, the only allowed single photon operator, up to possible additional factors of $ \partial^2$, is
\begin{eqnarray}
\int\! d^4\theta\; \Phi_1\Phi_2^\dagger \DD\WW_\vv+ \! \!\hc
\label{eqn:PhPhDW}
\end{eqnarray}
In the case of $\HH \leftrightarrow \HH$ interactions, the only operators of this form
allowed by the $O(2)_R$ and $U(1)_{\text{e+p}}$ symmetries
of the theory are 
\begin{eqnarray}
\int\!d^4\theta(\HH_1\HH_1^\dagger-\HH_2\HH_2^\dagger)\DD\WW_\vv \qquad \text{and} \qquad 
\int\!d^4\theta(\HH_1^c\HH_1^{c\dagger}-\HH_2^c\HH_2^{c\dagger})\DD\WW_\vv
\end{eqnarray}
These operators contain terms like $\varpi_+^\dagger\partial_\mu\varpi_+ \partial_\nu F^{\mu\nu}_\vv$ and
$\omega_+^\dagger\partial_\mu\omega_+ \partial_\nu F^{\mu\nu}_\vv$, respectively, and
thus correspond to charge radius interactions. Matching to Eq.~\ref{eqn:charge-radius-SUSY1} then gives the supersymmetric completion
of the charge radius interactions:
\begin{eqnarray}
\LL_{\HH \HH \vv} =  c_E g_\vv c_{2\theta} R_e^2 \int\!\!d^4 \theta  (\HH^\dagger_1 \HH_1 - \HH^{\dagger}_2 \HH_2 -\HH^{c}_1\HH_1^{c\dagger} + \HH^{c\dagger }_2 \HH^c_2)\DD \WW_\vv
\label{eqn:HHinteraction}
\end{eqnarray}
Replacing $\DD\WW_V$ with the current $\JJ_\vv$ gives atom-ion scattering.  Similarly matching onto the Rayleigh scattering operator in Eq.~\ref{eqn:rayleigh-SUSY} yields 
\begin{align}
\LL_{\HH\HH\vv\vv}= c_{E}^{\prime} g_\vv R_e^3 &\int\!d^4\theta\;  (\HH_1\HH_1^c+\HH_2\HH_2^c)^\dagger \WW_\vv\WW_\vv +\!\!\hc
\label{eqn:HHVVinteraction}
\end{align}
Just like the operator in Eq.~\ref{eqn:HHinteraction}, this operator will mediate decays within the hypermultiplet once supersymmetry is broken.

The restriction to operators of the form in Eq.~\ref{Eq: HV Int2} is a supersymmetric analog of the statement that any interaction involving two scalars and a  field strength can be written as ``$(\textup{derivatives})\times \phi\partial_\mu\phi'\partial_\nu F^{\mu\nu}$,''  which implies that transitions between scalar states cannot proceed via single photon emission. 
Thus, for example, direct single photon/photino decays from the $2s$ hypermultiplet to the ground state hypermultiplet are forbidden. The decay will instead proceed through either two photon/photino transitions or a cascade decay via magnetic operators of the form in Eq.~\ref{Eq: HV Int1}.

\subsubsection*{Hypermultiplet Decays}

In the supersymmetric limit, the hypermultiplet is exactly stable.  Once supersymmetry is broken and decays
within the hypermultiplet become kinematically allowed, it is interesting to ask what decay channels determine
the relaxation timescale.  This question
is complicated by the fact that supersymmetry breaking enters the physics of decays in a number of ways.  On the
one hand, supersymmetry breaking perturbs eigenvalues and eigenstates; this opens up phase space, changes the
equations of motion, and induces decay channels through mixing.  
On the other hand, supersymmetry breaking perturbs the effective interactions of the non-relativistic constituents.
The rest of this section considers these possibilities in more detail,
with the conclusion that eigenstate mixing in the magnetic spin-flip operator, Eq.~\ref{Eq: HV Int1},
induces the largest decay rates.

In the presence of soft masses, the supersymmetric operators in Eq.~\ref{eqn:HHinteraction} and \ref{eqn:HHVVinteraction} 
can mediate decays within the hypermultiplet.  In the case of the three-body decays mediated by the Rayleigh scattering operator in 
Eq.~\ref{eqn:HHVVinteraction}, these soft masses appear in the eight powers of phase space:
\begin{eqnarray}
\Gamma_{\{ \omega_{1+},\; \omega_{2+},\;\varpi_-\}\rightarrow\xi_\pm \Lambda_\vv \vv} \simeq \Gamma_{\xi_\pm\rightarrow\omega_- \Lambda_\vv \vv}
&\simeq& \nonumber
|c^{\prime}_E g_\vv R_e^3|^2 \frac{(m_{\xi_\pm}-m_{\omega_-})^8}{ 64 \pi^3 m_\HH} \\ &=&|c^{\prime}_E|^2 \frac{\alpha_\vv^{27}}{16 \pi^2}     \left(\frac{s_{2\theta}}{2}\right)^{18}   \left(\frac{m_{\text{soft}}}{m_{\text{FS}}}\right)^8  \mu
\end{eqnarray}
In the case of the two-body decays mediated by the charge radius operator in Eq.~\ref{eqn:HHinteraction},
these soft masses appear in phase space as well as in an overall factor of $m_{\tilde{\vv}}^2$.  This latter
factor arises from the modified equations of motion for $\Lambda_\vv$, which imply that
$\DD\WW_V \supset  \bar\theta \slashed{\partial} \Lambda_\vv \propto m_{\tilde{\vv}} \bar\theta \Lambda_\vv$.
The resulting decay rate is
\begin{eqnarray}
\nonumber
\Gamma_{\{ \omega_{1+},\; \omega_{2+},\;\varpi_-\}\rightarrow\xi_\pm \Lambda_\vv } \simeq 
\Gamma_{\xi_\pm\rightarrow\omega_- \Lambda_\vv } &\simeq&  |c_E g_\vv c_{2\theta} R_e^2|^2 \;|m_{\tilde{\vv}}(m_\HH-m_{\xi_\pm})|^2 \frac{(m_{\xi_{\pm}}-m_{\omega_-})^2}{4\pi m_\HH}\\
&\simeq&   |c_E|^2 c_{2\theta}^2  \frac{\alpha_\vv^{21}}{16}  \left(\frac{s_{2\theta}}{2}\right)^{6}  \left(\frac{m_{\text{soft}}}{m_{\text{FS}}}\right)^4  \left(\frac{m_{\tilde{\vv}}}{m_{\text{FS}}}\right)^2 \mu.
\label{eqn:HHV-rate}
\end{eqnarray}
Here two powers of $m_{\text{soft}}$ arise from cancellations between the terms involving $\HH$ and $\HH^c$ in Eq.~\ref{eqn:HHinteraction}. 
Higher order operators may not have this cancellation.

Next consider how the magnetic spin-flip operator in Eq.~\ref{Eq: HV Int1} induces decays in
the presence of supersymmetry breaking.  Mixing between, e.g., the fermionic states $\chi_\pm$ and $\xi_\pm$ allows all the states in 
$\HH$ to decay down to $\omega_-$
through Eq.~\ref{Eq: HV Int1}, which contains interactions of the form
\begin{eqnarray}
c_M g_\vv R_m s_{2\theta} m_{\tilde{\vv}}  \bar\Lambda_\vv  \begin{pmatrix}{1}\\ { \gamma_5 } \end{pmatrix}  \chi \begin{pmatrix}{\varpi}\\ { \omega } \end{pmatrix}^{\!\dagger}
\label{eqn:magneticchi}
\end{eqnarray}
Comparison with Eq.~\ref{eqn:fermionmatrix} shows that, in Eq.~\ref{eqn:magneticchi}, this fermionic mixing is 
accounted for by making a replacement of the form 
\begin{eqnarray}
\chi \rightarrow \chi +\frac{s_{4\theta}}{4} \frac{m_{\text{soft}}}{m_{\text{FS}}} \xi
\end{eqnarray}
which leads to the decay rate
\begin{eqnarray}
\Gamma_{\{ \omega_{1+},\; \omega_{2+},\;\varpi_-\}\rightarrow\xi_\pm \lambda_\vv} &\simeq& \Gamma_{\xi_\pm\rightarrow\omega_- \lambda_\vv} \nonumber
\simeq |c_M g_\vv s_{2\theta} R_m m_\vv|^2   \left( \frac{s_{4\theta}}{4} \frac{m_{\text{soft}}}{m_{\text{FS}}} \right)^{\!2} \frac{ (\tfrac{1}{2} s_{2\theta}^2 m_{\text{soft}})^2}{4\pi m_\HH }  \\
   &=& |c_M|^2 \alpha_\vv^9 \left( \frac{s_{2\theta} s_{4\theta}}{8} \right)^{\!2}  \left( \frac{m_{\text{soft}}}{m_{\text{FS}}} \right)^{\!4} \mu\;  
\label{eqn:gammamixie}
\end{eqnarray}
This decay rate also receives contributions from supersymmetry breaking in the effective Yukawa
operators of the non-relativistic theory, since the coefficients carry factors of
$m_e^{-1/2}$ and $m_p^{-1/2}$ from the non-relativistic normalization of the scalar
constituents in Eq.\;\ref{eqn:non-relfields}.  These contributions, however, are parametrically smaller
by an amount $\OO(m_{\text{FS}}^2/m_e^2)$.
Hence the decay rate Eq.\;\ref{eqn:gammamixie}, which is suppressed by only four powers of the largest 
supersymmetry breaking spurion, $m_{\text{soft}}$,
characterizes the relaxation timescale of the hypermultiplet. 

\subsection{$U(1)_\aa$ interactions }
\label{Sec: U1a}
This section outlines the dominant interactions between dark atoms and an axial $U(1)$ with charges given in Eq.\;\ref{Eq: Axial Charges} and mediated by a vector superfield $\aa$.   Axial gauge symmetry forbids mass terms for fermions and therefore the gauge symmetry must be broken if non-relativistic bound states exist. 
As in the previous section, there are several allowed supersymmetric operators
and the interactions of the vector boson are sufficient to fix the coefficients of the operators.

The leading $\aa_\mu$ interactions are determined by the axial charges of the constituents.  
The scalars $\varpi_\pm$ in Eq.~\ref{eqn:wavefunctions} have zero axial charge, but the combinations
\begin{equation}
\label{eqn:AxialInteraction0}
\frac{1}{\sqrt{2}}\left(\omega_+ \mp c_{2\theta}\omega_-\pm s_{2\theta}\varsigma_-\right)
\end{equation}
have charges of $\pm 2$ respectively. This leads to the inelastic interactions
\begin{equation}
2ig_\aa\left(c_{2\theta}\omega_+^\dagger\overleftrightarrow{\partial_\mu}\omega_- -s_{2\theta}\omega_+^\dagger\overleftrightarrow{\partial_\mu}\varsigma_-\right)\aa^\mu.
\label{eqn:AxialInteraction1}
\end{equation}
Similarly, the fermion-fermion bound states are charged with interactions given by
\begin{eqnarray}
g_\aa m_\vv \left(c_{2\theta}\varsigma_-^\dagger+s_{2\theta}\omega_-^\dagger\right)v_\mu\aa^\mu.
\label{eqn:AxialInteraction2}
\end{eqnarray}
The interactions of Eq.\;\ref{eqn:AxialInteraction0}-\ref{eqn:AxialInteraction2} can be embedded 
in the following superspace operators:
\begin{eqnarray}
&\LL_{\HH\HH\aa}\propto& g_\aa c_{2\theta}\int\!\!d^4 \theta\; \left(\HH_1^{c\dagger} \HH_1^c + \HH_2^{c\dagger}\HH_2^c\right)\aa
\label{eqn:AxialHyper}\\
&\LL_{\HH\VV\aa} \propto& g_\aa s_{2\theta}m_\vv\int\!\!d^4 \theta\; (\HH_1^c-\HH_2^{c\dagger})\VV \aa+\hc\\
&\LL_{\VV\VV\aa}\propto& g_\aa m_\vv^2 c_{2\theta}\int\!\!d^4 \theta\; \VV^\dagger\VV\aa.
\end{eqnarray}
The interactions of the linear superfield eaten by $\aa$ can be obtained by going out of unitary gauge
\begin{eqnarray}
\aa \rightarrow \aa + \frac{ \pi_\aa + \pi^\dagger_\aa}{\sqrt{2} m_\aa} .
\end{eqnarray}

In models like that of Sec.\;\ref{Sec: Nearly Susy}, where $\aa_\mu$ undergoes kinetic mixing with Standard Model
hypercharge, $\aa_\mu$ interactions mediate the dominant coupling of dark atoms to the Standard Model.
This setup can be used to realize inelastic dark matter because the elastic interaction of the ground state, $\omega_-$, with $\aa_\mu$ is forbidden due to parity.
After supersymmetry breaking, 
$\tfrac{1}{\sqrt 2}(\omega_+\pm\varpi_+)$ become mass eigenstates.  The interactions in 
Eq.\;\ref{eqn:AxialInteraction1}-\ref{eqn:AxialInteraction2} then allow $\omega_-$ to upscatter to 
$\omega_{1+}$ and $\omega_{2+}$,
which are heavier by an amount $\sim \OO(m_{\text{soft}})$, and to $v_\mu$, which is heavier by
an amount $\sim \OO(m_{\text{FS}})$.  

Higher dimension operators also contribute to the interactions with the standard model. For example, the axial spin flip operator
\begin{eqnarray}
\LL_{\HH\VV\aa}^{D=5}\propto  g_\aa s_{2\theta}R_m \int\!\!d^2 \theta\; \left(\HH_1^c\WW_1-\HH_2^c\WW_2\right)\WW_\aa+\textup{h.c.}
\end{eqnarray}
leads to scattering which can be important in certain regions of parameter space.

Two body decays mediated by $\aa_\mu$ are either kinematically
forbidden or severely suppressed, since $m_\aa$ can only be made smaller than $m_{FS}$ by 
choosing $g_\aa \lesssim \OO(\alpha_\vv^4)$.  Similarly, three body decays mediated by an off-shell
$\aa_\mu$ are subdominant.

\section{Kinetically Mixed Supersymmetric Hydrogen}
\label{Sec: Nearly Susy}

This section constructs a minimal model for a nearly supersymmetric dark sector that supports Coulombic bound states.  Sec.~\ref{Sec: Kinetic Mixing} introduces a minimal Higgs sector and discusses how kinetic mixing of the dark $U(1)_\aa$ with hypercharge in the supersymmetric Standard Model (SSM) drives gauge symmetry breaking in the hidden sector.  Sec.~\ref{Sec: Charged Matter} adds matter fields that are charged under a second Abelian gauge symmetry, $U(1)_\vv$, that introduces hydrogen-like bound states into the spectrum of the theory.  In the low energy limit this theory reduces to supersymmetric QED with two massive flavors.
Sec.~\ref{Sec: SSB} discusses how supersymmetry breaking is communicated to the dark sector from the SSM.
Finally, Sec.~\ref{Sec: Benchmarks} illustrates the scales of the resulting model by calculating three benchmark points.

\subsection{Kinetic Mixing}
\label{Sec: Kinetic Mixing}

Abelian field strengths are gauge invariant and therefore no symmetry principle forbids mixed field strength terms \cite{Holdom:1985ag}.  
Kinetic mixing occurs in extensions of the Standard Model with additional $U(1)$ gauge factors if there are fields that are charged under both the new $U(1)$ and hypercharge.     
In supersymmetric theories, the entire gauge supermultiplet undergoes gauge kinetic mixing, leading to both gaugino kinetic mixing and $D$-term mixing
\cite{SusyKineticMixing,Cheung:2009}.  
If there are light fields charged under the new $U(1)$, then kinetic mixing drives gauge symmetry breaking.

Consider a minimal example where a dark $U(1)_\aa$ couples to a pair of chiral superfields $\Phi$ and $\Phi^c$ with charges $\pm 2$ (chosen for later convenience).
The Lagrangian is given by
\begin{eqnarray}
\label{Eq: Simple L}
\!\!\!\!\!\!\LL_{\text{Hidden}} = \int\!\!d^4 \theta \;  (\Phi^\dagger e^{4 g_\aa \aa} \Phi + \Phi^c{}^\dagger e^{-4 g_\aa \aa} \Phi^c)
+\!\! \int\!\!d^2\theta\;(\tfrac{1}{4} \WW_\aa^2 -\tfrac{\epsilon}{2} \WW_\aa \WW_Y + W_0)+\!\!\hc
\end{eqnarray}
where $\aa$ is the supersymmetric gauge potential of the hidden $U(1)_\aa$, $\WW_\aa$ is the supersymmetric gauge field strength of $\aa$, and $\WW_Y$ is the supersymmetric gauge field strength of SSM hypercharge.
For $\epsilon \ll 1$ the hidden sector is only a small perturbation to the SSM so that all SSM fields will have their normal vacuum
expectation values; in particular the SSM Higgs fields will acquire vevs along a non $D$-flat direction:
\begin{eqnarray}
D_Y  = \frac{g_Y v^2}{4} \cos 2\beta 
\end{eqnarray}
This SSM vev now acts as a source term for $\WW_\aa$ in Eq.~\ref{Eq: Simple L}
and forces $\phi$, the lowest component of  $\Phi$, to acquire a vev, since $D_Y$ acts an effective Fayet-Illiopoulos term for $U(1)_\aa$.   The resulting effective Lagrangian is
\begin{eqnarray}
\LL_{D} = -\half D_\aa^2 + D_\aa( \epsilon D_Y  -2g_\aa( |\phi|^2 -|\phi^c|^2) ) \;\;\;   \Rightarrow  \;\;\;
|\phi|^2 = |\phi^c|^2 + \frac{ \epsilon D_Y}{ 2g_\aa}  \ne 0
\end{eqnarray}
This $D$-term potential has a residual flat direction, which can be lifted by $W_0$.   This section uses a superpotential 
\begin{eqnarray}
W_0=\lambda S (\Phi \Phi^c -\mu_0^2)
\label{eqn:W0}
\end{eqnarray}
where $S$ is a new singlet chiral superfield. 
With the addition of $W_0$ the vevs of all fields are fixed and there are no massless fermions.  It is convenient to let the superfields acquire vevs and to expand around the new field origin
\begin{eqnarray}
\langle \Phi \rangle = v_\aa \cos \beta_\aa \qquad \langle \Phi^c\rangle = v_\aa \sin \beta_\aa \qquad \langle \Phi \Phi^c \rangle = \mu_0^2 
\label{eqn:vevs}
\end{eqnarray}
where the last expression is enforced by the $F$-term for $S$.  
Solving for $v_\aa$ and $\tan\beta_\aa$ gives
\begin{eqnarray}
v^4_\aa= \left( \frac{\epsilon D_Y}{2g_\aa}\right)^2 + 4 \mu_0^4\qquad
\tan 2\beta_\aa = \frac{ 4 g_\aa \mu_0^2}{\epsilon D_Y}.
\label{eqn:va}
\end{eqnarray}
In the limit $\mu_0^2\ll \epsilon D_Y/g_\aa$, $\tan\beta_\aa \rightarrow 0$ and in the opposite limit, $\tan \beta_\aa \rightarrow 1$.
Fluctuations around the vacuum in Eq.~\ref{eqn:vevs} can be diagonalized using the field definitions
\begin{eqnarray}
\Phi =   ( v_\aa + \pi)\cos\beta_\aa + \sin\beta_\aa S^c \qquad   \Phi^c =( v_\aa - \pi)\sin\beta_\aa + \cos\beta_\aa S^c
\end{eqnarray}
so that the superpotential becomes
\begin{eqnarray}
W_0 = \lambda S( \Phi \Phi^c  -\mu_0^2) = \lambda v_\aa S S^c + \cdots 
\end{eqnarray}
clearly showing that $S$ picks up a Dirac mass  $m_S= \lambda v_\aa$.  In the K\"ahler term, the super-Higgs mechanism takes place:
\begin{eqnarray}
\nonumber
K &=&  \Phi^\dagger e^{4 g_\aa \aa}\Phi + \Phi^c{}^\dagger e^{-4g_\aa \aa} \Phi^c + S^\dagger S\\
&=& 
S^c{}^\dagger S^c + S^\dagger S+  m_{\aa}^2\left( \aa +  \frac{ \pi + \pi^\dagger}{\sqrt{2} m_\aa}\right)^2   +\cdots
\end{eqnarray}
The superfield $\pi$ is clearly identified as the eaten linear superfield, and the vector field has picked up a mass 
\begin{eqnarray}
m_\aa =  2\sqrt{2} g_\aa v_\aa.  
\end{eqnarray}

In addition to driving $U(1)_\aa$ gauge symmetry breaking, kinetic mixing also leads to $U(1)_R$-breaking mass effects in the gaugino sector of the theory.  This is because for $\epsilon \ne0$ gaugino kinetic mixing between $U(1)_Y$ and $U(1)_\aa$ entangles the
 $U(1)_R$-preserving Dirac mass $m_\aa$ with the $U(1)_R$-breaking bino mass $M_1$.  The effective Lagrangian for the gauginos is
\begin{eqnarray}
\!\!\LL_{ \lambda} =
\bar{\lambda}_Y i \sld \lambda_Y 
+ \bar{\lambda}_\aa i \sld \lambda_\aa
+\bar{\chi}_\pi i \sld \chi_\pi 
-( M_1 \lambda_Y \lambda_Y
+\epsilon \bar{\lambda}_\aa \sld \lambda_Y + m_\aa \chi_\pi \lambda_\aa +\!\!\hc\!\!)
\end{eqnarray}
where $\chi_\pi$ is the fermion component of the linear superfield $\pi$.
The three eigenvalues to $\OO(\epsilon^2)$ are
\begin{eqnarray}
m = m_\aa \left( 1 - \frac{ \epsilon^2 m_\aa}{M_1 - m_\aa}\right),\quad
- m_\aa \left( 1 - \frac{ \epsilon^2 m_\aa}{M_1 + m_\aa}\right),\quad
M_1 \left( 1 - \frac{ \epsilon^2 M_1^2}{m_1^2 -m^2_\aa}\right)\!\!.\quad
\end{eqnarray}
Notice that the two mass eigenvalues at $|m|\simeq m_\aa$ are no longer identically the same due to kinetic mixing with $\lambda_Y$, and this introduces $U(1)_R$ breaking into the hidden sector.

\subsection{Charged Matter}
\label{Sec: Charged Matter}

This section adds light, charged matter to the dark sector.
 The charged matter consists of four chiral superfields $E$, $E^c$, $P$, and $P^c$ that have axial charges under $U(1)_\aa$. 
The charge assignments of the dark electron and proton are chosen to be chiral to prevent them
from acquiring supersymmetric masses in the
absence of gauge symmetry breaking.  Once the $U(1)_\aa$ gauge symmetry
is broken,  these states acquire masses at a scale set by $m_\aa$ and $\tan \beta_\aa$.    
In addition to new matter fields, the gauge sector is extended by a second gauge group $U(1)_\vv$ under which $E$, $E^c$, $P$, and $P^c$ have
vector-like charges and which will lead to the formation of hydrogen-like bound states in the hidden sector.  
In summary the additional matter content has the following charge assignments:
\begin{eqnarray}
\begin{array}{|c|cccc|}
\hline
& E& E^c& P & P^c\\
\hline
U(1)_\vv& -1& +1 & + 1& -1\\
U(1)_\aa& +1 & +1 & -1 & -1\\
\hline
\end{array} 
\end{eqnarray}
The superpotential in Eq.~\ref{eqn:W0} is augmented by Yukawa terms
\begin{eqnarray}
\label{eqn:Wfull}
W = W_0 + W_{\text{Yukawa}} \qquad\qquad W_{\text{Yukawa}}  = y_e \Phi^c E E^c + y_p \Phi P P^c  +  \!\!\hc
\end{eqnarray}
so that after $U(1)_\aa$ breaking both $E$ and $P$ acquire Dirac masses
\begin{eqnarray}
\!\!\!\!W_{\text{Yukawa}} = m_e\!\left( \! 1 + \frac{-\pi  + \cot\beta_\aa  S^c}{v_\aa}\right) \!E E^c +m_p\!\left(\!1 + \frac{\pi  + \tan\beta_\aa  S^c}{v_\aa}\right)\!P P^c + \!\!\hc
\end{eqnarray}
where
\begin{eqnarray}
m_e = y_e \sin \beta_\aa v_\aa \qquad \text{and} \qquad m_p = y_p \cos \beta_\aa v_\aa
\end{eqnarray}
None of the fields charged under $U(1)_\vv$ acquires a vev, and therefore $U(1)_\vv$ is a massless gauge multiplet.

The interactions of the $U(1)_\aa$ vector superfield $\aa$ with the matter superfields are given by 
\begin{eqnarray}
K = 2g_{\aa}\left( \aa + \frac{\pi + \pi^\dagger}{\sqrt{2}m_\aa}\right)  (E^\dagger E + E^c{}^\dagger E^c - P^\dagger P - P^c{}^\dagger P^c).
\end{eqnarray}
Here the interactions of the $\pi$ have been moved from the superpotential to the K\"ahler potential 
with the equations of motion.   The $\pi$ fields can have subdominant mixing with the Higgs fields of the SSM and
mediate subdominant interactions. 

The Higgs trilinear coupling, $\lambda$, in Eq.~\ref{eqn:W0} and the axial gauge coupling, $g_\aa$, 
are taken to be large enough that the masses in the axial/Higgs sector are of order $\alpha_\vv m_e$ or larger.  
With this choice of parameters the axial and Higgs sectors decouple, and
the low energy limit of the theory is supersymmetric QED with two massive flavors and weakly broken supersymmetry.
Because the axial/Higgs sectors respect the $O(2)_R$ symmetry, the arguments of Sec.~\ref{Sec: Wavefunctions Fields}
go through and, in particular, the leading order superspin wavefunctions are as given in Sec.~\ref{eqn:wavefunctions}.
The dominant residual effect of the axial/Higgs sector is to perturb the mass splitting between the hypermultiplet
and vector multiplet.  These contributions are suppressed through a combination of coupling constants 
and/or Yukawa suppression.   
Although the axial $U(1)_\aa$ gauge sector plays
a subdominant role in the internal dynamics of the hidden sector, it mediates the dominant coupling to the 
Standard Model.  In particular it mediates supersymmetry breaking, which is the subject of the next section.

\subsection{Supersymmetry Breaking}
\label{Sec: SSB}
\declareslashed{}{/}{0.1}{0}{R}
Although the hidden sector is supersymmetric at tree level, at the loop level small supersymmetry breaking
effects are induced through the kinetic mixing portal to the SSM. 
This section discusses the strength with which the constituent particles' masses feel supersymmetry breaking.
These soft masses determine the leading supersymmetry breaking effects in the ground state spectrum, as discussed
in Sec.~\ref{Sec: SSB Effects}.

The soft parameters to be calculated (see Eq.~\ref{eqn:lpreserve} and Eq.~\ref{eqn:lbreak}) 
are the $U(1)_R$-preserving $\Delta^2_{\tilde{e}}$ and $\Delta^2_{\tilde{p}}$ and the $U(1)_R$-breaking 
$B_e$ and $B_p$.  The largest soft parameters are the $U(1)_R$-preserving ones.
If supersymmetry breaking is mediated to the SSM through gauge mediation, then these are given by
\begin{eqnarray}
\Delta^2_{\tilde{e} } \simeq \Delta^2_{\tilde{p}} \simeq \frac{\alpha_\aa\epsilon^2}{\alpha'} M^2_{\tilde{E}^c}
\end{eqnarray}
where $M_{\tilde{E}^c}$ is the SSM right handed selectron mass.
The next largest soft parameters are the $U(1)_R$-breaking $B_{\mu}$-type terms, $B_e$ and $B_p$.  To
isolate how $U(1)_R$-breaking effects are mediated from the SSM, it is useful to integrate out the bino, 
which generates the operator
\begin{eqnarray}
O_{\slashed{R}} = \lambda_\aa \frac{\epsilon^2 M_1 \square}{ \square + M_1^2} \lambda_\aa +\!\! \hc
\end{eqnarray}
$B_e$ and $B_p$ are then generated upon insertion of $O_{\slashed{R}}$ in a loop, with
a logarithmically enhanced contribution that is the same for both, 
\begin{eqnarray}
B_e \simeq B_p \simeq B \equiv \frac{\alpha_{\aa}\epsilon^2 M_1}{\pi} \text{log} \frac{\Lambda_{UV}^2}{M_1^2} 
\end{eqnarray}
where $\Lambda_{UV}$ is the messenger scale.
The $B$-terms feed into the $U(1)_\vv$ gaugino mass, which is highly suppressed
due to the indirect communication of $U(1)_R$-breaking
\begin{eqnarray}
\label{Eq: Vector gaugino mass}
m_{\tilde{\vv}} \sim \frac{  \alpha_\vv B}{4\pi}   \sim \frac{\alpha_\aa \alpha_\vv \epsilon^2}{(4 \pi)^2}M_1.
\end{eqnarray}
In this model, $\lambda_\vv$ is always light and its mass is smaller than the level splittings induced by supersymmetry breaking,
which are of order $m_{\text{soft}}$
\begin{eqnarray}
\frac{m_{\tilde{\vv}}}{m_{\text{soft}}} \sim \frac{\alpha_\vv \alpha' }{(4\pi)^2}\frac{  M_1 m_e }{M_{\tilde{E}^c}^2} \ll 1. 
\end{eqnarray}
This justifies ignoring the contributions from $m_{\tilde{\vv}}$ to the ground state energy levels.  

Sec.\;\ref{Sec: SSB Effects} described how the dominant communication of supersymmetry breaking to the spectrum is
through supersymmetry violating perturbations to the rest energies of the constituents.
 Supersymmetry breaking also introduces several dynamical contributions to bound state spectroscopy from
 the exchange of particles frfom the axial $U(1)_\aa$ and Higgs sectors.    However, these contributions are suppressed
 for the same reason that the supersymmetric contributions from the axial and Higgs sectors are suppressed.

\subsection{Benchmark Models}
\label{Sec: Benchmarks}
This section constructs three benchmark models to illustrate the  scales that emerge in the hidden sector.  Although doing detailed direct detection phenomenology is outside the scope of this paper, in both cases
we aim to construct spectra compatible with iDM phenomenology.  In particular we require that the bound states have
a mass $m_\text{DM} \sim 100 \GeV$ and a splitting $\delta \sim 100 \keV$ between the ground state and the next 
highest state accessible through axial scattering.  Furthermore the (predominantly inelastic) scattering cross section between the ground state
and standard model nucleons should be $\OO(10^{-40}) \;\text{cm}^2$.

It is possible to meet these criteria; however, some tension exists between meeting all three criteria
simultaneously.  In models consistent with these requirements $\omega_-$ or $\varsigma_-$ is the lightest state.  The three states to which $\omega_-$/$\varsigma_-$ can upscatter by exchanging an axial photon with a nucleon are $\omega_{1+}$, $\omega_{2+}$, and $v_\mu$. The cross section for $\omega_-$/$\sigma_-$ to upscatter to $v_\mu$ is velocity suppressed and can be ignored in the following. The cross section for $\omega_-$ to upscatter to $\omega_{1+}$ or $\omega_{2+}$ via the operator in Eq.\;\ref{eqn:AxialInteraction1} is given\cite{Cheung:2009,Alves:2009nf} by
\begin{eqnarray}
\sigma \simeq 64 \pi \frac{ \alpha_\text{EM} \alpha_\aa \epsilon^2 m_\text{N}^2}{m_\aa^4}
=\frac{\alpha_{\text{EM}}m_\text{N}^2}{D_Y^2}\frac{1}{1+\frac{16g_\aa^2\mu_0^4}{\epsilon^2D_Y^2}}
\simeq \frac{4 \times10^{-41}\text{ cm}^2}{1+\frac{16g_\aa^2\mu_0^4}{\epsilon^2D_Y^2}}\frac{(50 \GeV)^4}{D_Y^2}
\end{eqnarray}
where $m_\text{N}$ is the mass of the nucleon.  In order to fix $\sigma \sim10^{-40} \text{ cm}^2$, one must therefore choose $\mu_0^2\lesssim\epsilon D_Y/g_\aa$.  For natural values of the Yukawa couplings, $m_\text{DM}\lesssim v_\aa$, and so from Eq.\;\ref{eqn:va} and the requirement that
$m_\text{DM} \sim 100 \GeV$ follows the constraint that
\begin{eqnarray}
\frac{\epsilon}{g_\aa}\gtrsim\frac{(100\GeV)^2}{D_Y}.
\end{eqnarray}
Bounds on kinetic mixing \cite{Hook:2010tw} impose further constraints, requiring $\epsilon\lesssim0.005$ for $m_\aa \sim 1 \GeV$. Finally, the scale of supersymmetry breaking is proportional to $\epsilon^2 g_\aa^2$ and to get splittings of order $100\keV$ requires that $\epsilon\sim g_\aa\sim0.005$. 

Trying to match the CoGeNT/DAMA anomaly \cite{Essig:2010ye} with light inelastic dark matter is challenging in this specific model because it is
difficult to generate an $\OO(10^{-38}\text{ cm}^2)$ cross section.   The primary tension arises from the mediation through the
massive axial current.   One approach could be to kinetically mix the vector current rather than the axial current.  This does not
suppress elastic scattering; however, in models of inelastic dark matter where the signal arises from downscattering, rather than
upscattering, the elastic rate does not need to be suppressed.  
This possibility is not pursued further
here but illustrates the rich phenomenology possible in composite dark matter models \cite{Lisanti:2009am}.

Following the above logic we choose the following MSSM parameters for all three benchmark models
\begin{eqnarray}
\begin{array}{|c|c|c|c|}
\hline
 \sqrt{D_Y} & M_1&M_{\tilde{E}^c}&\Lambda_\text{UV}\\
\hline
 50\GeV & 100 \GeV&900\GeV&100\TeV\\
\hline
\end{array} .
\end{eqnarray}
The parameters of the dark sector for the three benchmark points are chosen to be
\begin{eqnarray}
\begin{array}{|c||c|c|c|c|c|c|}
\hline
\text{Model}& \epsilon & g_\vv & g_\aa &  \mu_0 & y_e & y_p\\
\hline
\text{Unmixed}&0.005& 1.5 & 0.004 &  30 \GeV & 0.15 & 1.2\\
\hline
\text{Mixed}&0.005& 1.1 & 0.004 &  30 \GeV & 0.25 & 1.4 \\
\hline
\text{Heavy Scalars}&0.005&0.5&0.004&30\GeV&0.35&1.0\\
\hline
\end{array} .
\end{eqnarray}
The first two choices for $g_\vv$ cause $U(1)_\vv$ to hit a Landau pole before the GUT scale.  
The Landau pole can be avoided by embedding $U(1)_\vv$ into a non-Abelian group, 
{\em e.g. } $U(1)_\vv \subset SU(2)_\vv$ or $U(1)_\vv\times U(1)_\aa \subset SO(4)$. The
rather large values of $\epsilon$ chosen here make for some tension with constraints from BaBar.  These
constraints may not apply to this model because $\aa_\mu$ may cascade decay through the Higgs sector  
before decaying into Standard Model particles \cite{ExperSignatures}.
These parameters lead to the following supersymmetric bound state mass scales
\begin{eqnarray}
\begin{array}{|c||c|c|c|c|c|c|c|}
\hline
\text{Model}& v_\aa &m_\aa& m_e & m_p & \tan\theta&m_{\text{Prin}}& m_{\text{FS}} \\
\hline
\text{Unmixed}&49 \GeV& 550\MeV& 3.0 \GeV& 53\GeV&0.24 &46\MeV& 1500\keV\\
\hline
\text{Mixed}&49 \GeV& 550\MeV& 5.1\GeV& 62\GeV&0.29 &22\MeV& 200\keV\\
\hline
\text{Heavy Scalars}&49\GeV&550\MeV&7.1\GeV&44\GeV&0.40&1.2\MeV&0.48\keV\\
\hline
\end{array}
\end{eqnarray}
Supersymmetry breaking effects are encapsulated in the soft parameters
\begin{eqnarray}
\begin{array}{|c||c|c|c|}
\hline
\text{Model} &m_\text{soft}& B & m_{\tilde{\vv}} \\
\hline
\text{Unmixed}& 270\keV& 6.1\eV & 0.09 \eV  \\
\hline
\text{Mixed}& 170 \keV & 6.1 \eV & 0.05\eV \\
\hline
\text{Heavy Scalars}&130\keV&6.1\eV&0.01\eV\\
\hline
\end{array} .
\end{eqnarray}
The gauge-mediated contribution to $m_{\tilde{\vv}}$ listed here is subdominant to gravity-mediated contributions, 
which give
\begin{eqnarray}
m_{\tilde{\vv}} \sim \frac{F_{\text{susy}}}{M_\text{pl}} \sim 1 \eV 
\end{eqnarray}
where $\sqrt{F_{\text{susy}}} \simeq 100 \TeV$ for these benchmark models.

\FIGURE[t]{
\centering
\qquad\includegraphics[width=6.20in]{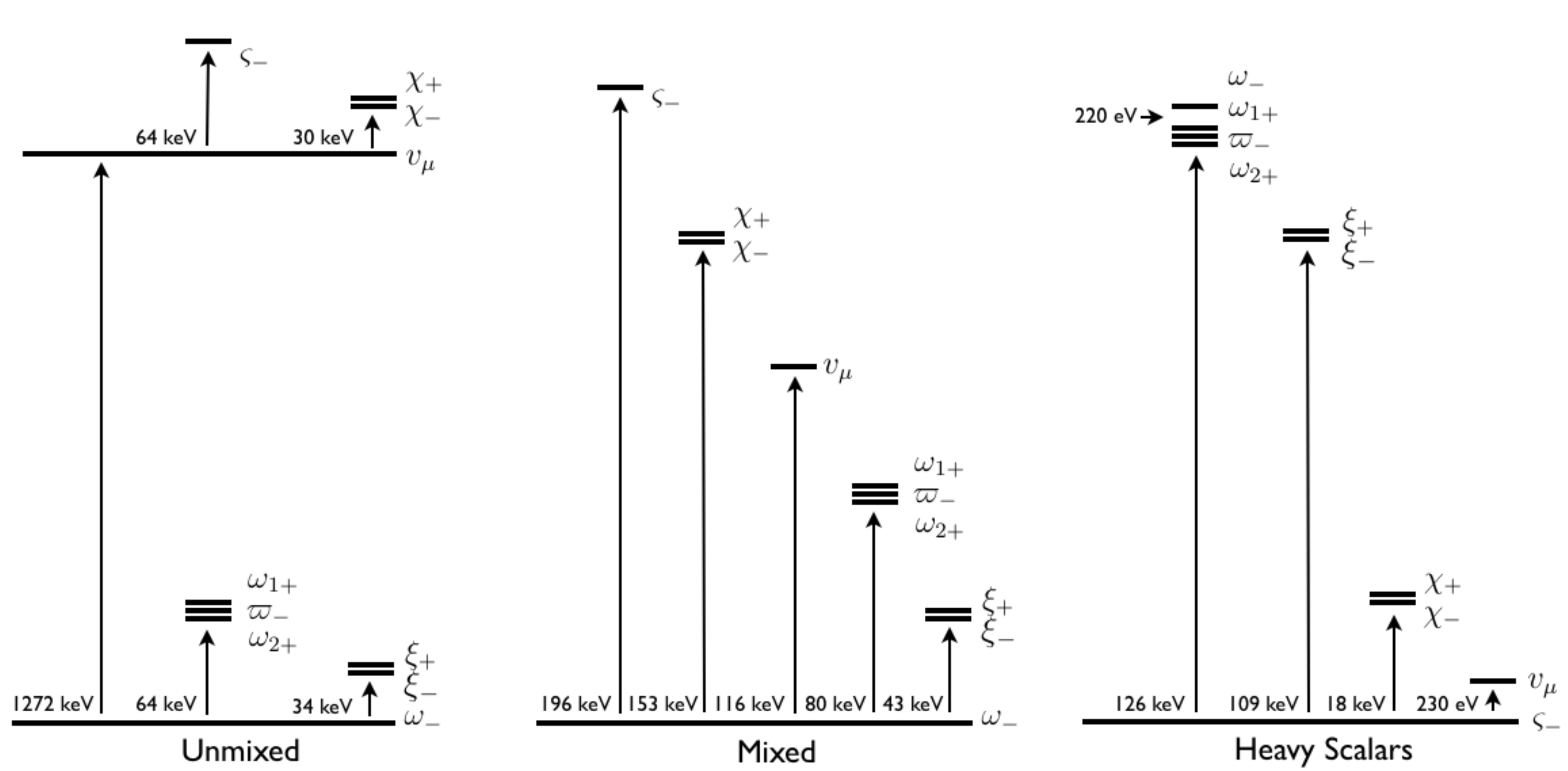}\qquad
\caption{The ground state spectrum of the three benchmark models.  In all cases the small, unlabeled 
splittings are of order $B \simeq 6 \eV$.}}

The resulting spectra are shown in Figure 2.  The first parameter point, ``Unmixed," realizes
the scenario where $B \ll m_\text{soft}\ll m_{FS}$.  The various bound states
have a mass $ 56 \GeV$ and the lowest state $\omega_-$ is primarily scalar-scalar.  The lowest
states accessible by axial photon exchange are the pair of nearly degenerate scalars, $\omega_{1+}$ and $\omega_{2+}$,
which are heavier by an amount $\delta = 64 \keV$.
For the second parameter point, ``Mixed," the hierarchy of scales is instead 
$B \ll m_\text{soft}\sim m_\text{FS}$ and there is large mixing between the vector multiplet and hypermultiplet. The bound
states have a mass $67 \GeV$ and an
iDM-compatible spectrum is again realized with several states available for upscattering.  $\omega_-$ is again the lightest
state but the vector $v_\mu$ is now kinematically accessible.  Because scattering to $v_\mu$ is velocity
suppressed, however, the relevant splitting for iDM is $\delta =80 \keV$ between $\omega_-$ and  $\omega_{1+}$/$\omega_{2+}$.
For the third parameter point, ``Heavy Scalars," the hierarchy of scales is instead 
$B \ll m_\text{FS}\ll m_\text{soft}$ and $m_\text{DM} = 51 \GeV$.  
Because the hypermultiplet has large selectron components, the hypermultiplet
states move upwards and the spectrum is inverted, with $\varsigma_-$ (which is primarily fermion-fermion) and 
$v_\mu$ as the lightest states.  The relevant  splitting for iDM is $\delta = 126 \keV$.

From the results in Sec.\;\ref{Sec: U1v} one can estimate the lifetime of the unstable states in the ``Unmixed" benchmark
spectrum.  One finds $\tau_\VV \sim 10^{-18} \text{ sec}$ for the vector multiplet and 
$\tau_\HH \sim 10^{-13} \text{ sec}$ for the states in the hypermultiplet.  Although the decay formulae from Sec.\;\ref{Sec: U1v}
are not directly applicable to the ``Mixed" benchmark, the decay rates will be similarly fast, since $\alpha_\vv$ is $\OO(1)$.

\section{Discussion }
\label{Sec: Discussion} 
The benchmark models discussed in Sec.\;\ref{Sec: Benchmarks} demonstrate that the non-relativistic
supersymmetric bound states discussed in this article can realize a wide range of bound state spectra with
a rich hierarchy of scales.
In any application of these models to dark matter phenomenology, 
a number of important issues must be addressed.  In particular one needs to examine constraints
from direct and indirect detection as well as the implications for early universe cosmology.  Although
addressing these topics in any detail lies outside of the scope of this paper, in this section we briefly
discuss some of the relevant physics.  In the following our discussion is limited to the concrete model
presented in Sec.\;\ref{Sec: Nearly Susy}.

\subsection{BBN Constraints }

In a realistic model where the dark matter sector is nearly supersymmetric,
the hidden $U(1)_\vv$ photon and photino that create the bound states will usually be relativistic at
the temperatures relevant for BBN, $T \approx 1$ MeV.  This means that they contribute to the energy density of the universe
and will modify the successful predictions of standard BBN.
Fortunately, the resulting perturbation is sufficiently small to be within observational constraints.
This arises because the hidden sector is weakly coupled to the visible sector,  typically $\epsilon \lesssim 10^{-2}$,
and therefore the dark sector kinetically decouples \cite{Bringmann:2006mu} from the visible sector at a temperature above the $\GeV$ scale.  The early kinetic decoupling reduces the number of effective degrees of freedom in the dark sector because many Standard Model degrees
of freedom become non-relativistic between temperatures of 10 GeV and the QCD phase transition.  This generally makes 
 weakly coupled dark sectors with long range Coulombic interactions \cite{BuckleyFox} safe from BBN constraints \cite{Ackerman:2008gi}.

\subsection{Recombination} 

If dark atoms are relevant for cosmology, then a significant fraction of the supersymmetric dark electrons and protons must recombine into supersymmetric atoms.
The recombination of non-supersymmetric hydrogen-like dark matter atoms is studied in  \cite{Kaplan:2009de}. Supersymmetry adds new levels of complexity to the problem. 
First, supersymmetry introduces new processes where gauginos are emitted in de-excitation processes.  
Sec.~\ref{Sec: Interactions} discusses some of these de-excitation processes and, as a rule of thumb, gaugino emission processes are faster if the corresponding gauge boson emission process is a magnetic, spin-flip transition.  
For non-supersymmetric dark atoms, the most important processes for recombination are the Ly-$\alpha$ decay, the $2s$ double photon decay and the scattering process $e^-+p^+\leftrightarrow H+\gamma$ \cite{Kaplan:2009de}. None of these are due to spin-flips and so gaugino emission processes 
are expected to be subdominant.

Even if gaugino emission is subleading as expected, there are new electric transitions between the different, non-degenerate superspin levels with $n=2$ and $n=1$.   This additional complexity makes the out of equilibrium problem harder to solve systematically.  The numerous new Ly-$\alpha$ lines corresponding to different superspin transitions are sufficiently degenerate that the Doppler broadening smears the energies of the states.   For this reason photons from different transitions are indistinguishable implying that the optical depth of Ly-$\alpha$ photons is about the same as in the non-supersymmetric case.
Although a detailed analysis is necessary, it appears that the physics of recombination in the supersymmetric case is parametrically the same as 
in the non-supersymmetric case.  In particular for sufficiently large $\alpha_\vv$ recombination is expected to be an efficient process.

\subsection{Molecules}

If supersymmetric atoms can form, it is possible that these atoms may further aggregate into supersymmetric molecules.
This section briefly explores this possibility by examining the role that Bose/Fermi statistics plays in atoms and molecules.

The ground state of regular diatomic hydrogen is a $s=0$ state, i.e.\;the spin wavefunction for the electrons is antisymmetric under particle exchange and the spatial wavefunction is symmetric as it would be for a two-selectron atom. The statistics of the electrons thus does not affect the ground state spatial wavefunction and energy to lowest order. Supersymmetric hydrogen should therefore form diatomic molecules, and the binding energy is approximately the same as for non-supersymmetric hydrogenic systems\cite{Clavelli:2008zs}.

In the Standard Model, further aggregation into molecules larger than $H_2$ is prevented by the
Pauli exclusion principle, which forbids more than two electrons from being in the same orbital. 
In supersymmetric atoms, this aggregation is not forbidden by Pauli because electrons can convert into their scalar superpartners, selectrons.  Supersymmetric bound states will share orbitals more effectively and hence are bound more strongly.  
For non-relativistic molecules composed of $N$ bosonic constituents, the binding energy scales as
\begin{eqnarray}
|E_{\text{binding}}|\propto N^{\frac{7}{5}},
\label{eqn:BindingEnergyBound}
\end{eqnarray}
in contrast to molecules with fermionic constituents, whose binding energy grows linearly with the number of constituents \cite{Dyson,Conlon:1987xt,Lieb:1988tt}. This means that macroscopic bound states formed from scalar constituents have an enormous binding energy. This suggests a scenario where a fraction of the dark atoms condense into huge ``molecules," which may impose additional constraints on supersymmetric atomic dark matter if the formation of macro-molecules is too efficient.   One possibility is that the supersymmetric macro-molecules could drive formation of microscopic black holes.  

\subsection{Dark Matter Genesis}

In order for dark atoms to be a sizeable fraction of the Universe's dark matter, there needs to be a chemical potential for $U(1)_{e+p}$ generated in the early Universe \cite{Sakharov}.   The most compelling mechanism for generating an asymmetry in the dark matter sectors links the dark matter number density to the baryonic or leptonic number density, for recent work see references in \cite{Darkogenesis}.
The primary novelty with generating a chemical potential for the composite sector is that the minimal gauge invariant operator sourcing $U(1)_{e+p}$ is a dimension 2 operator,  {\em e.g.} $E P$.  This typically means that constructing a renormalizable Lagrangian requires more structure than for elementary dark matter.  
Normal inelastic dark matter requires $n_b/n_{e+p} \gsim 1$ and this means that if the $e+p$ asymmetry is to be directly linked to the baryon asymmetry, then the interactions that equilibrate  the chemical potentials in the dark sector and Standard Model must freeze out when the dark matter is relativistic.  Alternatively, the $e+p$ asymmetry should be generated through a  mechanism like the out-of-equilibrium decay of a heavy particle.

\section{Conclusions }
\label{Sec: Conc}

Composite dark matter offers a rich phenomenology that has only barely been explored in comparison to models
in which dark matter is an elementary particle.  Non-relativistic bound states offer one general class of composite 
dark matter models, and these
typically involve a new mass scale that is incorporated either by hand or through a Higgs mechanism.  In the later case the Higgs mass is radiatively unstable and supersymmetry is a natural way to stabilize its mass.   Supersymmetry breaking can be weakly communicated to this sector, particularly if the interactions of the dark sector with the Standard Model are the dominant link to the supersymmetry breaking sector.   If this is the case, then the composite dark matter will form nearly supersymmetric multiplets and the phenomenology of these states can be radically different from the non-supersymmetric case.

Atomic Inelastic Dark Matter \cite{Kaplan:2009de} was proposed as a model of inelastic dark matter where the hyperfine splitting of the ground state of a hydrogen-like sector is the origin of the
 inelastic mass splitting.  This article created
 a supersymmetric version of Atomic Inelastic Dark Matter that has many different features from the original, non-supersymmetric model.   Most notably, the hyperfine splitting disappears in the 
 supersymmetric limit and the ground state typically contains a scalar-scalar component.    This introduces new interactions between dark matter and the Standard Model.  The model accommodates spectra and scattering cross sections compatible with iDM phenomenology.    
 
 This article has also constructed tools to help in the study of quasi-perturbative 
 supersymmetric bound states and in incorporating supersymmetry breaking into the bound states.   
 These tools illustrate that the form of the weakly broken spectrum and the composition of the various states is in many 
 cases dictated entirely by supersymmetry at leading order.  Supersymmetry also imposes strict restrictions on
 the allowed interactions, which simplifies the matching of effective interaction operators.
  More generally supersymmetric bound states offer a rich laboratory for studying 
 supersymmetric dynamics and interactions.  

\section*{Acknowledgements}
JGW thanks E. Katz for useful conversations during the course of this work.
MJ and TR would like to acknowledge helpful conversations with Michael Peskin
as well as collaboration with Daniele Alves in early stages of this work.
TR is a William R. and Sara Hart Kimball Stanford Graduate Fellow.
  MJ, SB and JGW  are supported by the US DOE under contract number DE-AC02-76SF00515.  
MJ, SB, TR and JGW receive partial support from the Stanford Institute for Theoretical Physics.  
JGW is partially supported by the US DOE's Outstanding Junior Investigator Award and the Sloan Foundation.  
JGW thanks the Galileo Galilei Institute for their hospitality during the early stages of this work.

\providecommand{\href}[2]{#2}\begingroup\raggedright

\endgroup

\end{document}